\def\powerMax{P_{\rm max}}
\def\pathLossExpo{\alpha}

\def\deltaVectorAtIteration[#1][#2]{{\Delta}^{(#1)}_{#2}}
\def\deltaVectorAtIterationEle[#1]{{\bm \Delta}^{#1}}

\def\signNormal[#1]{\sign\left(#1\right)}
\def\numbOfTrainingImages{D}
\def\complexNumbers{\mathbb{C}}
\def\realNumbers{\mathbb{R}}

\def\constante{{\rm e}}
\def\constantj{{\rm j}}

\def\expectationOperator[#1][#2]{{\mathbb{E}_{#2}}\left[#1\right]}
\def\indicatorFunction[#1]{\mathbb{I}\left[{#1}\right]}
\def\probability[#1]{\mathbb{P}\left[{#1}\right]}

\def\numberOfEdgeDevices{K}
\def\timeDomainOFDM[#1]{x(#1)}
\def\timeDomainOFDMNorm[#1]{x_{\rm norm}(#1)}
\def\timeVar{t}

\def\indexSubcarrier{l}
\def\numberOfActiveSubcarriers{M}
\def\idftSize{N}
\def\dataSymbols[#1]{d_{#1}}

\def\symbolDuration{T_{\rm s}}

\def\receivedSymbolAtSubcarrier[#1]{r_{#1}}
\def\transmittedSymbolAtSubcarrier[#1]{t_{#1}}
\def\randomSymbolAtSubcarrier[#1]{s_{#1}}
\def\channelAtSubcarrier[#1]{h_{#1}}
\def\noiseAtSubcarrier[#1]{n_{#1}}
\def\numberOfOFDMSymbols{S}
\def\indexOFDMSymbol{m}
\def\asymbolFromED[#1]{d_{#1}}

\def\exponentialIntegral[#1]{{\rm Ei}(#1)}
\def\tciFactor[#1]{p_{#1}}
\def\mappingFunction{{f}}
\def\encoder[#1]{\psi_#1}
\def\symbolEnergy{E_{\rm s}}

\def\voteInTime[#1]{m^{#1}}
\def\voteInFrequency[#1]{l^{#1}}

\def\noiseVariance{\sigma_{\rm n}^2}

\def\coefficientOne{a}

\def\correctDecision[#1]{p_{#1}}
\def\incorrectDecision[#1]{q_{#1}}
\def\aparameterForBer[#1]{\epsilon_{#1}}

\def\probabilityIncorrect[#1]{P^{\rm err}_{#1}}

\def\effectiveSNR{\xi}

\def\identiyVector[#1]{\textbf{\textrm{I}}_{#1}}
\def\zeroVector[#1]{\textbf{\textrm{0}}_{#1}}

\def\dataset[#1]{\mathcal{D}_{#1}}

\def\datasetBatch[#1]{\mathcal{\tilde{D}}_{#1}}
\def\batchSize{n_{\rm b}}
\def\brachSizeRelativeToRounds{\gamma}
\def\completeData{\mathcal{D}}
\def\numberOfModelParameters{q}
\def\sampleData[#1]{{\textrm{\textbf{x}}}_{#1}}
\def\sampleLabel[#1]{{y}_{#1}}
\def\learningRate{\eta}

\def\indexED{k}
\def\indexGradient{i}

\def\indexCommunicationRound{n}

\def\modelParametersAtIteration[#1]{\textbf{w}^{(#1)}}
\def\modelParametersAtIterationEle[#1][#2]{w^{(#1)}_{#2}}

\def\modelParameters{\textbf{w}}
\def\modelParametersEle[#1]{{w}_{#1}}

\def\localGradientSign[#1][#2]{\bar{\textbf{g}}_{#1}^{(#2)}}
\def\localGradient[#1][#2]{\tilde{\textbf{g}}_{#1}^{(#2)}}
\def\localGradientNoIndex[#1]{\tilde{\textbf{g}}_{#1}}

\def\localGradientSignElement[#1][#2]{{\bar{g}}_{#1}^{(#2)}}
\def\localGradientElement[#1][#2]{{\tilde{g}}_{#1}^{(#2)}}
\def\localGradientNoIndexElement[#1]{{\tilde{g}}_{#1}}

\def\lossFunctionSample[#1]{f(#1)}
\def\lossFunctionLocal[#1][#2]{F_{#1}(#2)}
\def\lossFunctionGlobal[#1]{F(#1)}
\def\lossFunctionGlobalMinimum{F^*}

\def\majorityVoteEle[#1][#2]{{v}^{(#1)}_{#2}}
\def\majorityVote[#1]{\textbf{v}^{(#1)}}

\def\globalGradient[#1]{{\textbf{{g}}}^{(#1)}}
\def\globalGradientElement[#1][#2]{{{g}}^{(#1)}_{#2}}
\def\globalGradientNoIndex{{\textbf{{g}}}}
\def\globalGradientElementNoIndex[#1]{{g_{#1}}}

\def\communicationRounds{N}

\def\metricForFirst[#1]{r_{#1}^{+}}
\def\metricForSecond[#1]{r_{#1}^{-}}

\def\nonnegativeConstants{\textbf{L}}
\def\nonnegativeConstantsEle[#1]{L_{#1}}
\def\varianceBound{{\bm \sigma}}
\def\varianceBoundEle[#1]{\sigma_{#1}}

\def\rmsDelaySpread{T_{\rm rms}}

\def\symbolVector[#1]{\textbf{\textrm{d}}_{#1}^{(\indexCommunicationRound)}}
\def\symbolVectorEstimate[#1]{\tilde{\textbf{\textrm{{d}}}}_{#1}}
\def\receivedVector[#1]{\textbf{\textrm{r}}_{#1}}
\def\noiseVector[#1]{\textbf{\textrm{n}}_{#1}}

\def\transmittedVector[#1]{\textbf{\textrm{t}}_{#1}}
\def\idftMatrix[#1]{\textbf{\textrm{F}}_{#1}^{\rm H}}
\def\dftMatrix[#1]{\textbf{\textrm{F}}_{#1}}
\def\transformPrecoder[#1]{\textbf{\textrm{D}}_{#1}}
\def\transformDecoder[#1]{\textbf{\textrm{D}}_{#1}^{\rm H}}
\def\frequencyMapping{\textbf{\textrm{M}}_{\textrm{f}}}

\def\channelMatrix[#1]{\textbf{\textrm{H}}_{#1}}

\def\rawCubicMetric[#1]{{\rm RCM}_{#1}}
\def\slopeFactor{K_{\rm s}}
\def\rawCubicMetricRef{{\rm RCM}_{\rm ref}}
\def\cubicMetric[#1]{{\rm CM}(#1)}

\def\syncError{T_{\rm sync}}
\def\channelSpread{T_{\rm chn}}

\def\guardTimeUniform{T_{\rm g}}
\def\numberOfVotesPerDFTsOFDM{M_{\rm v}}
\def\uniformGap{M_{\rm g}}

\def\diagonalization[#1]{{\rm diag}\{#1\}}

\def\referenceDistance{R_{\rm ref}}
\def\minimumDistance{R_{\rm min}}
\def\cellRadius{R_{\rm max}}
\def\distanceED[#1]{r_{#1}}
\def\powerED[#1]{P_{#1}}
\def\pathlossExponent{\alpha}
\def\powerControl{\beta}

\def\transmittedSignalDiscrete[#1]{p\left[#1\right]}
\def\transmittedSignal[#1]{p\left(#1\right)}
\def\receivedSignalDiscrete[#1]{r\left[#1\right]}
\def\receivedSignal[#1]{r\left(#1\right)}
\def\receivedSignalDiscreteFrequency[#1]{b_{#1}}
\def\numberOfOccupiedSubcarriers{D}

\def\numberOfShifts{M}
\def\indexSubcarrier{j}

\def\basisFunction[#1]{B_{#1}(\timeVar)}
\def\amountOfShift[#1]{\tau_{#1}}
\def\dataSymbols[#1]{d_{#1}}
\def\angleSignal[#1]{\psi_{#1}(\timeVar)}
\def\instantaneousFrequency[#1]{F_{#1}(t)}
\def\besselFunctionFirstKind[#1][#2]{J_{#1}\left(#2\right)}
\def\lowerFrequency{L_{\rm d}}
\def\upperFrequency{L_{\rm u}}
\def\idftSize{N}

\def\fourierSeries[#1]{c_{#1}}

\def\timeVar{t}

\def\fresnelC[#1]{C(#1)}
\def\fresnelS[#1]{S(#1)}
\def\linearXone{\alpha_\indexSubcarrier}
\def\linearXtwo{\beta_\indexSubcarrier}
\def\linearCoef{\gamma_\indexSubcarrier}

\newcommand\mydots{\hbox to 1em{.\hss.\hss.}}

\documentclass[conference,comsoc]{IEEEtran}
\usepackage{multicol}
\usepackage{lipsum}
\usepackage{mathtools}
\usepackage{amsthm}
\usepackage{acronym}
\usepackage{stfloats}
\usepackage{amsfonts}
\usepackage{acronym}
\usepackage{cite}
\usepackage{multirow}
\usepackage{bm}
\usepackage{amsmath,amssymb}
\usepackage{graphicx}
\usepackage[table,xcdraw]{xcolor}
\usepackage[caption=false,font=footnotesize]{subfig}
\usepackage[geometry]{ifsym}
\usepackage{array}
\usepackage[utf8]{inputenc}
\usepackage[T1]{fontenc}
\let\norm\undefined % <-- "Undefine" \norm
\DeclarePairedDelimiter\norm{\lVert}{\rVert}

\usepackage{tikz}
\usepackage{lipsum}
\usetikzlibrary{calc,shadings,patterns}
\makeatletter
\tikzset{%
  remember picture with id/.style={%
    remember picture,
    overlay,
    save picture id=#1,
  },
  save picture id/.code={%
    \edef\pgf@temp{#1}%
    \immediate\write\pgfutil@auxout{%
      \noexpand\savepointas{\pgf@temp}{\pgfpictureid}}%
  },
  if picture id/.code args={#1#2#3}{%
    \@ifundefined{save@pt@#1}{%
      \pgfkeysalso{#3}%
    }{
      \pgfkeysalso{#2}%
    }
  }
}

\def\savepointas#1#2{%
  \expandafter\gdef\csname save@pt@#1\endcsname{#2}%
}

\def\tmk@labeldef#1,#2\@nil{%
  \def\tmk@label{#1}%
  \def\tmk@def{#2}%
}

\tikzdeclarecoordinatesystem{pic}{%
  \pgfutil@in@,{#1}%
  \ifpgfutil@in@%
    \tmk@labeldef#1\@nil
  \else
    \tmk@labeldef#1,(0pt,0pt)\@nil
  \fi
  \@ifundefined{save@pt@\tmk@label}{%
    \tikz@scan@one@point\pgfutil@firstofone\tmk@def
  }{%
  \pgfsys@getposition{\csname save@pt@\tmk@label\endcsname}\save@orig@pic%
  \pgfsys@getposition{\pgfpictureid}\save@this@pic%
  \pgf@process{\pgfpointorigin\save@this@pic}%
  \pgf@xa=\pgf@x
  \pgf@ya=\pgf@y
  \pgf@process{\pgfpointorigin\save@orig@pic}%
  \advance\pgf@x by -\pgf@xa
  \advance\pgf@y by -\pgf@ya
  }%
}

\makeatother
\newcounter{hatchNumber}
\DeclarePairedDelimiter\ceil{\lceil}{\rceil}
\DeclarePairedDelimiter\floor{\lfloor}{\rfloor}

\usepackage{cuted}
\makeatletter
\newif\ifAC@uppercase@first%
\def\Aclp#1{\AC@uppercase@firsttrue\aclp{#1}\AC@uppercase@firstfalse}%
\def\AC@aclp#1{%
	\ifcsname fn@#1@PL\endcsname%
	\ifAC@uppercase@first%
	\expandafter\expandafter\expandafter\MakeUppercase\csname fn@#1@PL\endcsname%
	\else%
	\csname fn@#1@PL\endcsname%
	\fi%
	\else%
	\AC@acl{#1}s%
	\fi%
}%
\def\Acp#1{\AC@uppercase@firsttrue\acp{#1}\AC@uppercase@firstfalse}%
\def\AC@acp#1{%
	\ifcsname fn@#1@PL\endcsname%
	\ifAC@uppercase@first%
	\expandafter\expandafter\expandafter\MakeUppercase\csname fn@#1@PL\endcsname%
	\else%
	\csname fn@#1@PL\endcsname%
	\fi%
	\else%
	\AC@ac{#1}s%
	\fi%
}%
\def\Acfp#1{\AC@uppercase@firsttrue\acfp{#1}\AC@uppercase@firstfalse}%
\def\AC@acfp#1{%
	\ifcsname fn@#1@PL\endcsname%
	\ifAC@uppercase@first%
	\expandafter\expandafter\expandafter\MakeUppercase\csname fn@#1@PL\endcsname%
	\else%
	\csname fn@#1@PL\endcsname%
	\fi%
	\else%
	\AC@acf{#1}s%
	\fi%
}%
\def\Acsp#1{\AC@uppercase@firsttrue\acsp{#1}\AC@uppercase@firstfalse}%
\def\AC@acsp#1{%
	\ifcsname fn@#1@PL\endcsname%
	\ifAC@uppercase@first%
	\expandafter\expandafter\expandafter\MakeUppercase\csname fn@#1@PL\endcsname%
	\else%
	\csname fn@#1@PL\endcsname%
	\fi%
	\else%
	\AC@acs{#1}s%
	\fi%
}%
\edef\AC@uppercase@write{\string\ifAC@uppercase@first\string\expandafter\string\MakeUppercase\string\fi\space}%
\def\AC@acrodef#1[#2]#3{%
	\@bsphack%
	\protected@write\@auxout{}{%
		\string\newacro{#1}[#2]{\AC@uppercase@write #3}%
	}\@esphack%
}%
\def\Acl#1{\AC@uppercase@firsttrue\acl{#1}\AC@uppercase@firstfalse}
\def\Acf#1{\AC@uppercase@firsttrue\acf{#1}\AC@uppercase@firstfalse}
\def\Ac#1{\AC@uppercase@firsttrue\ac{#1}\AC@uppercase@firstfalse}
\def\Acs#1{\AC@uppercase@firsttrue\acs{#1}\AC@uppercase@firstfalse}

\newtheorem{theorem}{Theorem}

 \newtheorem{assumption}{Assumption}
\DeclareMathOperator{\sign}{sign}

\acrodef{W-CDMA}{Wideband Code Division
Multiple Access}
\acrodef{ACLR}{adjacent-channel-leakage ratio}
\acrodef{OBO}{output-power back-off}
\acrodef{OOB}{Out-of-band emission}
\acrodef{signSGD}{sign stochastic gradient descent}
\acrodef{CSC}{circularly-shifted chirp}
\acrodef{SNR}{signal-to-noise ratio}
\acrodef{RMSE}{root-mean-square error}
\acrodef{OFDM}{orthogonal frequency division multiplexing}
\acrodef{DFT}{discrete Fourier transform}
\acrodef{PSK}{phase-shift keying}
\acrodef{QAM}{quadrature amplitude modulation}
\acrodef{QPSK}{quadrature phase-shift keying}
\acrodef{PMEPR}{peak-to-mean envelope power ratio}
\acrodef{CM}{cubic metric}
\acrodef{RCM}{raw cubic metric}
\acrodef{BER}{bit-error ratio}
\acrodef{SNR}{signal-to-noise ratio}
\acrodef{PSD}{power spectral density}
\acrodef{SE}{spectral efficiency}
\acrodef{CP}{cyclic prefix}
\acrodef{AWGN}{additive white Gaussian noise}
\acrodef{CFR}{channel frequency response}
\acrodef{CIR}{channel impulse response}
\acrodef{MMSE}{minimum mean square error}
\acrodef{LMMSE}{linear minimum mean square error}
\acrodef{BPSK}{binary phase shift keying}
\acrodef{BLER}{block-error rate}
\acrodef{ML}{maximum likelihood}
\acrodef{PHY}{physical layer}
\acrodef{PA}{power amplifier}
\acrodef{IDFT}{inverse DFT}
\acrodef{DoF}{degrees-of-freedom}
\acrodef{IoT}{Internet-of-Things}
\acrodef{FDE}{frequency-domain equalization}
\acrodef{RF}{radio-frequency}
\acrodef{IM}{index modulation}
\acrodef{BS}{base station}
\acrodef{MF}{matched filter}
\acrodef{PPM}{pulse-position modulation}

\acrodef{BAA}{broadband analog aggregation}
\acrodef{OBDA}{one-bit broadband digital aggregation}
\acrodef{FEEL}{federated edge learning}
\acrodef{FL}{federated learning}
\acrodef{ED}{edge device}
\acrodef{ES}{edge server}
\acrodef{UL}{uplink}
\acrodef{DL}{downlink}
\acrodef{OAC}[OAC]{over-the-air computation}
\acrodef{TCI}{truncated-channel inversion}
\acrodef{MV}{majority vote}
\acrodef{CNN}{convolution neural network}
\acrodef{ReLU}{rectified-linear unit}
\acrodef{CSI}{channel state information}
\acrodef{PAPR}{peak-to-average power ratio}
\acrodef{SC}{single-carrier}
\acrodef{iid}[IID]{independent and identically distributed}
\acrodef{RMS}{root-mean-square}
\acrodef{4G}{Fourth Generation}
\acrodef{5G}{Fifth Generation}
\acrodef{NR}{New Radio}
\acrodef{LTE}{Long-Term Evolution}
\acrodef{LoRa}{long-range}
\acrodef{FSK}{frequency-shift keying}
\acrodef{FSK-MV}{\ac{FSK}-based \ac{MV}}
\acrodef{PPM}{pulse-position modulation}
\acrodef{PPM-MV}{\ac{PPM}-based \ac{MV}}
\def\fdssVector{{\rm \bf f}}
\acrodef{CSC}{circularly-shifted chirp}
\acrodef{DFT-s-OFDM}{discrete Fourier transform-spread orthogonal frequency division multiplexing}
\acrodef{CSC-MV}{\ac{CSC}-based \ac{MV}}
\def\cellRadius{R_{\rm max}}

\def\distanceEDVar{R}
\def\oboVar{\textrm{OBO}}

\def\oboMinCsc{\textrm{OBO}_{\rm min}^{\rm csc\textrm{-}mv}}
\def\oboMin{\textrm{OBO}_{\rm min}}
\def\oboMax{\textrm{OBO}_{\rm {ref}}}
\def\oboMinObda{\textrm{OBO}_{\rm min}^{\rm obda}}
\def\rangePowercontrol{r_{\rm P}}

\def\figuresizeWidth{3.2in}
\acrodef{IID}[IID]{independent and identically}

\acrodef{LoRa}{Long Range}
\acrodef{HF}{high-frequency}
\acrodef{FDSS}{frequency-domain spectral shaping}
\acrodef{PUCCH}{physical uplink control channel}
\acrodef{OFDMA}{orthogonal frequency division multiple access}

\acrodef{TX}{transmitter}
\acrodef{RX}{receiver}
\acrodef{OAC}{over-the-air computation}

\def\BibTeX{{\rm B\kern-.05em{\sc i\kern-.025em b}\kern-.08em
    T\kern-.1667em\lower.7ex\hbox{E}\kern-.125emX}}
\begin{document}

\title{Chirp-Based Over-the-Air Computation for Long-Range Federated Edge Learning} 

% \author{
% 		\IEEEauthorblockN{Safi Shams Muhtasimul Hoque}
% 	\IEEEauthorblockA{Electrical  Engineering Department\\
% 	University of South Carolina\\
% 	Columbia, SC, USA\\
% 	Email: shoque@email.sc.edu}	
% 		\and	
% \IEEEauthorblockN{Alphan \c{S}ahin}
% 	\IEEEauthorblockA{Electrical  Engineering Department\\
% 	University of South Carolina\\
% 	Columbia, SC, USA\\
% 	Email: asahin@mailbox.sc.edu}
% }

\author{
	\IEEEauthorblockN{Safi Shams Muhtasimul Hoque, Mohammad Hassan Adeli, and Alphan \c{S}ahin} \IEEEauthorblockA{Electrical Engineering Department, University of South Carolina, Columbia, SC, USA\\
		Email: shoque@email.sc.edu, madeli@email.sc.edu, asahin@mailbox.sc.edu}
} 

% \author{
% \IEEEauthorblockN{Safi Shams Muhtasimul Hoque}
% 	\IEEEauthorblockA{Electrical  Engineering Department\\
% 	University of South Carolina\\
% 	Columbia, SC, USA\\
% 	Email: shoque@email.sc.edu}
% 	\and
% 	\IEEEauthorblockN{Alphan \c{S}ahin}
% 	\IEEEauthorblockA{Electrical  Engineering Department\\
% 	University of South Carolina\\
% 	Columbia, SC, USA\\
% 	Email: everetb@email.sc.edu}	
% 		\and	
% 	\IEEEauthorblockN{Mohammad Hassan Adeli}
% 	\IEEEauthorblockA{Electrical  Engineering Department\\
% 		University of South Carolina\\
% 		Columbia, SC, USA\\
% 		Email: madeli@email.sc.edu}
% }

\maketitle
\begin{abstract}

In this study, we propose \ac{CSC-MV}, a power-efficient \ac{OAC} scheme, to achieve long-range \ac{FEEL}. The proposed approach maps the votes (i.e., the sign of the local gradients) from the \acp{ED} to the linear \acp{CSC} constructed with a \ac{DFT-s-OFDM} transmitter. At the \ac{ES}, the \ac{MV} is calculated with an energy detector. We compare our proposed scheme with \ac{OBDA} and show that the \ac{OBO} requirement of the transmitters with an \ac{ACLR} constraint for \ac{CSC-MV} is lower than the one with \ac{OBDA}. For example, with an \ac{ACLR} constraint of $-22$~dB, \ac{CSC-MV} can have an \ac{OBO} requirement of $6-7$~dB less than the one with \ac{OBDA}. When the \ac{PA} non-linearity is considered, we demonstrate that \ac{CSC-MV} outperforms \ac{OBDA} in terms of test accuracy for both homogeneous and heterogeneous data distributions, without using \ac{CSI} at the \ac{ES} and \acp{ED}.

%, as well as how resource utilization affects the effectiveness of \ac{DFT-s-OFDM} in comparison to \ac{OBDA} with \ac{QAM}. 
\end{abstract}

\acresetall
\section{Introduction}
% 1) Description of FEEL
% 2) Why AirComp? What are the state of the art solutions?
\Ac{FEEL} is an implementation of \ac{FL} over a wireless network, in which many \acp{ED} participate in training and an \ac{ES} aggregates the local decisions without accessing the local data at the \acp{ED} \cite{gafni2021federated}. With \ac{FEEL}, a significant number of model parameters/gradients/updates needs to be exchanged between the \ac{ES} and the \acp{ED} through a band-limited wireless channel. However, in this scenario, conventional orthogonal multiple access techniques can cause a large per-round communication latency as the number of \acp{ED} grows. One solution to this problem is to exploit the signal superposition property of wireless channels to compute the necessary calculations needed for the training over the air \cite{chen2021distributed}. However, multipath fading, imperfect power control, and synchronization errors in a practical network can complicate the design of a reliable \ac{OAC} scheme. 
%Furthermore, the state-of-the-art \ac{OAC} schemes often rely on the availability of the \ac{CSI} at the \ac{ED} and \ac{ES}, which increases the complexity of the design. 
Also, several wireless communication metrics such as \ac{ACLR}, \ac{PMEPR}, \ac{CM}, and \ac{PA} efficiency need to be taken into account in practice as \ac{FEEL} heavily relies on the uplink transmission and the availability of a large number of \acp{ED} such as low-cost \ac{IoT} devices (e.g., battery-powered sensors) as in \ac{LoRa} networks. In this paper, we propose an \ac{OAC} method that particularly addresses the communication-related challenges of \ac{FEEL} with \ac{CSC} \cite{Sahin_CSC2020}.

% 3) Motivation for long range: IoT, Sensors. Long range means low PAPR
In the literature, \Ac{OAC}, which was initially considered for wireless sensor networks \cite{Goldenbaum_2013,Nazer_2007}, has recently been investigated for distributed learning over a wireless network with both digital and analog modulations. In \cite{Guangxu_2020}, amplitude modulation over \ac{OFDM} subcarriers, i.e., called \ac{BAA}, with the model parameters is proposed. To overcome the impact of multi-path fading on the aggregation, \ac{TCI} is utilized, where the symbols on the \ac{OFDM} subcarriers are multiplied by the inverse of the channel coefficients, and the fading subcarriers are not used. In \cite{Guangxu_2021}, \ac{OBDA}, where the signs of the gradients are mapped to the \ac{QPSK} symbols and transmitted along with \ac{TCI}, is investigated. In \cite{Amiria_2021}, the authors analyze a scenario where the \ac{ES} is equipped with multiple antennas and the \ac{CSI} is not available to the \acp{ED}. To achieve coherent combining, the \ac{ES} uses superposed \acp{CSI}.
In \cite{sahinCommnet_2021}, an \ac{OAC} scheme based on \ac{signSGD} \cite{Bernstein_2018} with \ac{MV} is proposed. This method does not rely on the \ac{CSI} since it calculates the \ac{MV} by comparing the energy accumulation over the orthogonal resources. Therefore, it is immune to time-synchronization errors. In the literature, there are few studies that consider the \ac{PMEPR} for \ac{OAC}. In \cite{sahinWCNC_2022}, it is shown that if the gradients are correlated, the resulting \ac{OFDM} symbol with \ac{OBDA} can cause very high instantaneous peak power. To address this issue, the signs of the gradients are represented as \ac{PPM} symbols constructed with \ac{DFT-s-OFDM}. Nevertheless, \ac{PMEPR} can still be high depending on the choice of the parameters, e.g., pulse duration in a \ac{PPM} symbol.

% 4) Your proposal
In this work, we propose a new \ac{OAC} approach based on \acp{CSC}, particularly tailored for scenarios where a power-efficient transmission is needed for \ac{FEEL} to achieve wider coverage. Our specific contributions are listed as follows:
\begin{itemize}
    \item By exploiting inherently low \ac{PMEPR} of \acp{CSC} \cite{Sahin_CSC2020} and inspired by the simplicity of distributed training by the \ac{MV} with \ac{signSGD} \cite{Bernstein_2018}, we show how to design chirp-based distributed learning over a wireless network. With the proposed scheme, we indicate the signs of local stochastic gradients by changing the positions of \acp{CSC}. We also do not use \ac{CSI} at the \acp{ED} and \ac{ES}. 
    \item We show that the proposed scheme requires less \ac{OBO} than that of \ac{OBDA} for a given \ac{ACLR} constraint. A reduced \ac{OBO} results in increased cell coverage, which enables long-range distributed learning in a wireless network.
    \item We demonstrate the proposed approach can provide high test accuracy under both homogeneous and heterogeneous data distribution scenarios. We show that the proposed \ac{OAC} scheme allows the model to learn the classes available at the cell-edge \ac{ED} by increasing the cell size under a spectral leakage constraint. 
\end{itemize}

% 5) Notation
{\em Notation:} The sets of complex and real numbers are denoted by $\complexNumbers$ and $\realNumbers$, respectively. 
%$\expectationOperator[\cdot][t]$ is the expectation of its argument over $t$. 
%The function $\sign(\cdot)$ results in $1$ and $-1$ for a positive and a negative argument, respectively. 
The function $\sign(\cdot)$ gives $1$ and $-1$ for a non-negative and a negative argument, respectively. The $N$-dimensional all zero and one vectors are $\zeroVector[{N}]$ and $\identiyVector[{N}]$, respectively. $\indicatorFunction[\cdot]$ denotes the indicator function. 
% $\sim\mathcal{CN}(\mu, \noiseVariance)$ denotes the complex normal distribution with mean $\mu$ and variance. $\noiseVariance$%We use the notation $(\textbf{a})_i^j$ as shorthand for denoting a vector $[{a}_i,{a}_{i+1},\mydots,{a}_j]^{\rm T}$. 

\section{System Model}
\label{sec:system}

\subsection{Deployment}
We consider a circular cell of radius $\cellRadius$~meters with an \ac{ES} at its center. We assume that all the \acp{ED} are deployed uniformly at a radial distance from $\minimumDistance$~meters to $\cellRadius$~meters from the \ac{ES}. All \acp{ED} and \ac{ES} are equipped with a single antenna. We consider time-synchronization errors while frequency synchronization is assumed to be perfect.

% We assume that the frequency synchronization and the time synchronization are also assumed to be ideal among the \acp{ED}. In practice, this assumption does not hold. Nevertheless,  to draw a fair comparison with \ac{OBDA}, which only works under perfect synchronization, we shall consider there are no synchronization errors.  Our proposed scheme relies on non-coherent computation as \cite{sahinCommnet_2021} and \cite{sahinWCNC_2022} demonstrate that the non-coherent \ac{OAC} schemes are robust to synchronization errors.

In this study, we consider a power control mechanism that considers the maximum transmit power constraint at the \acp{ED}. %In this mechanism, the transmit power of an \acp{ED} should be set to a desired level so that the path loss experienced by that \ac{ED} is compensated. 
To model this, let $\powerED[{\rm ref}]/\noiseVariance$ be the \ac{SNR} of an \ac{ED} at the \ac{ES} location when the corresponding link distance is $\referenceDistance$ meters. Without loss of generality, we consider $\powerED[\rm ref] = 1$~W. We then express the received signal power of the $\indexED$th \ac{ED} located at the distance $\distanceED[\indexED]$ away from the \ac{ES} as
\begin{align}
\powerED[{\distanceED[\indexED]}]=\begin{cases}
\left(\frac{{\distanceED[\indexED]}}{\referenceDistance}\right)^{-\pathlossExponent+\powerControl}\powerED[{\rm ref}], & 0\leq\distanceED[\indexED]<\rangePowercontrol\\
\left(\frac{{\rangePowercontrol}}{\referenceDistance}\right)^{-\pathlossExponent+\powerControl}\powerED[{\rm ref}], & \distanceED[\indexED]\geq\rangePowercontrol
\end{cases}~,
\end{align}
where $\pathlossExponent$ is the path loss exponent of the corresponding channel, $\powerControl\in [0,\pathlossExponent]$ is a coefficient that determines how much path loss is compensated, and $\rangePowercontrol>\referenceDistance$ is the threshold distance beyond which the \acp{ED} are unable to attain the desired \ac{SNR} at the \ac{ES}. %While $\powerControl=\pathlossExponent$ corresponds to a perfect power control, $\powerControl=0$ means a system without a power control mechanism.  
%The effective path loss exponent  $\effectivePathLossExponent$ is defined as $\effectivePathLossExponent\triangleq\pathlossExponent-\powerControl$. 
To determine $\rangePowercontrol$, we consider the impact of \ac{PA} non-linearity on the transmitted signals from \acp{ED} and set an \ac{ACLR} constraint as follows: Let $\oboMax$ be the \ac{OBO} for the link distance $\referenceDistance$ to achieve the desired \ac{SNR}. Also, let $\oboMin$ be the minimum \ac{OBO} (i.e., results in maximum spectral tolerable growth) that fulfills the \ac{ACLR} constraint. The path loss compensation can be maintained perfectly (i.e., $\powerControl=\pathlossExponent$) up to a range of $\rangePowercontrol$. Thus, $\powerED[{\distanceED[\indexED]}]=
1$ for $ 0\leq\distanceED[\indexED]<\rangePowercontrol$. The transmitted signal power of the \ac{ED} located at $\distanceED[\indexED]$ is $\left({{\distanceED[\indexED]}}/{\referenceDistance}\right)^{\powerControl}$ for $ 0\leq\distanceED[\indexED]<\rangePowercontrol$. Let $\powerMax$ be the maximum power output of the \ac{PA}. Then, $\oboMax=10\log_{10}\left(\powerMax\right/\powerED[{\rm ref}])$ and $\oboMin=10\log_{10}\left(\powerMax/\powerED[{\rm ref}]\left({\referenceDistance}/{{\rangePowercontrol}}\right)^{\powerControl}\right)$, we then calculate $\rangePowercontrol$ as 
% $\rangePowercontrol=\referenceDistance\times 10^{\frac{\oboMax-\oboMin}{10\powerControl}}$.
\begin{align}
    \rangePowercontrol=\referenceDistance\times 10^{\frac{\oboMax-\oboMin}{10\powerControl}}.
    \label{eq:rangeofpowercontrol}
\end{align} 
Therefore, for a signal with a lower \ac{PMEPR}, $\oboMin$ is a smaller value and the coverage is larger.

%The downlink power control is assumed to be perfect. We aim at analyzing the performance of our proposed scheme while taking the impact of power misalignment and the path loss on distributed learning into account.
\subsection{Signal Model: Circularly-Shifted Chirps}
\begin{figure*}[t]
	\centering
	%\vspace{0.5mm}
	{\includegraphics[width =6.7in]{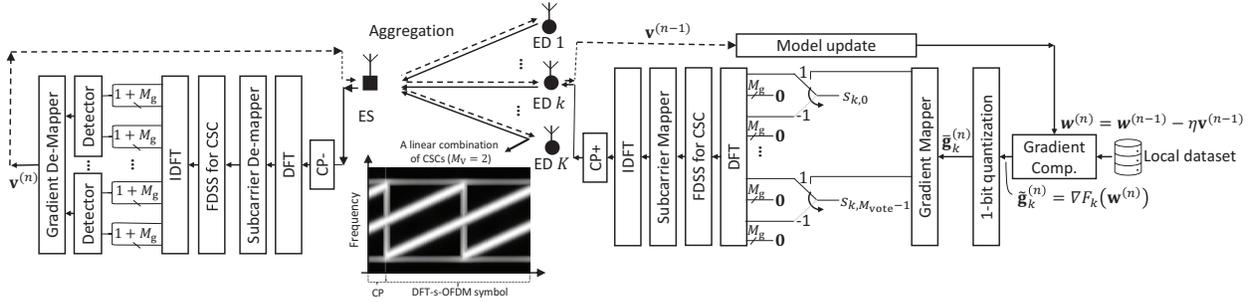}
	} 
	\caption{FEEL with \ac{CSC-MV}. Due to its compatibility to \ac{5G} \ac{NR}, DFT-s-OFDM with a special FDSS is utilized to generate multiple linear CSCs \cite{Sahin_CSC2020}.} 
	\label{fig:feelBlockDiagram}
\end{figure*}

\def\normalTerm{a_\textrm{N}}
In this work, we utilize the \ac{DFT-s-OFDM} block to carry the gradient information with chirps by using the method introduced in \cite{Sahin_CSC2020}. In \cite{Sahin_CSC2020}, it is shown that a chirp signal can be synthesized through \ac{DFT-s-OFDM} with a special choice of \ac{FDSS} coefficients with the motivation of its compatibility to 3GPP 4G LTE and 5G NR. In this study, we adopt this approach and the \acp{ED} access the spectrum with the \acp{CSC} synthesized through \ac{DFT-s-OFDM}, simultaneously, given by
\begin{align}
    \transmittedVector[\indexED,\indexOFDMSymbol] =\idftMatrix[\idftSize]\frequencyMapping\diagonalization[\fdssVector]\transformPrecoder[\numberOfActiveSubcarriers]\symbolVector[\indexED,\indexOFDMSymbol]~,
    \label{eq:txsymbol}
\end{align}
where $\transmittedVector[\indexED,\indexOFDMSymbol]\in \complexNumbers^\idftSize$ is the $\indexOFDMSymbol$th transmitted baseband signal in discrete 
time for the $\indexED$th \ac{ED}, $\idftMatrix[\idftSize]\in\complexNumbers^{\idftSize\times\idftSize}$ is the orthonormal $\idftSize$-point \ac{IDFT} matrix, $\transformPrecoder[\numberOfActiveSubcarriers]\in\complexNumbers^{\numberOfActiveSubcarriers\times\numberOfActiveSubcarriers}$ is the orthonormal $\numberOfActiveSubcarriers$-point \ac{DFT} matrix, $\frequencyMapping\in\realNumbers^{\idftSize\times\numberOfActiveSubcarriers}$ is the mapping matrix that maps the output of the \ac{DFT} precoder to a set of subcarriers, $\fdssVector\in\complexNumbers^\numberOfActiveSubcarriers$ is the \ac{FDSS} vector to synthesize chirps, and $\symbolVector[\indexED,\indexOFDMSymbol]\in\complexNumbers^{\numberOfActiveSubcarriers}$ contains the symbols on $\numberOfActiveSubcarriers$ bins. 
In \cite{Sahin_CSC2020}, 
it is shown that $\transmittedVector[\indexED,\indexOFDMSymbol]$ is a linear combination of {\em linear} \acp{CSC} where the amount of frequency sweep for each \ac{CSC} is $\frac{\numberOfOccupiedSubcarriers}{\symbolDuration}$ for symbol duration $\symbolDuration$ if the vector $\fdssVector$ is expressed as $\fdssVector=[\fourierSeries[\lowerFrequency],\dots,\fourierSeries[\upperFrequency]]^{\top}\times{\sqrt{\numberOfShifts/ {\sum_{\indexSubcarrier=\lowerFrequency}^{\upperFrequency}{|\fourierSeries[\indexSubcarrier]|^2}}}}$, where $\fourierSeries[\indexSubcarrier]$ is given by
\begin{align}
\fourierSeries[\indexSubcarrier] = \linearCoef(\fresnelC[{\linearXone}] + \fresnelC[{\linearXtwo}] + \constantj\fresnelS[{\linearXone}] + \constantj\fresnelS[{\linearXtwo}])~.
\label{eq:fresnekfcn}
\end{align}
$\fresnelC[{\cdot}]$ and $\fresnelS[{\cdot}]$ are the Fresnel integrals with cosine and sine functions, respectively, $\linearXone=(\numberOfOccupiedSubcarriers/2+2\pi\indexSubcarrier)/\sqrt{\pi\numberOfOccupiedSubcarriers }$, $\linearXtwo=(\numberOfOccupiedSubcarriers/2-2\pi\indexSubcarrier)/\sqrt{\pi\numberOfOccupiedSubcarriers}$, $\linearCoef= \sqrt{\frac{\pi}{\numberOfOccupiedSubcarriers}}\constante^{-\constantj\frac{(2\pi\indexSubcarrier)^2}{2\numberOfOccupiedSubcarriers}-\constantj\pi \indexSubcarrier}$ for $\indexSubcarrier\in\{\lowerFrequency,\mydots,\upperFrequency\}$, $\lowerFrequency\le -\numberOfOccupiedSubcarriers/2$, $\upperFrequency\ge \numberOfOccupiedSubcarriers/2$, and $\upperFrequency-\lowerFrequency+1=\numberOfActiveSubcarriers$. The spectrogram of a linear combination of two linear \acp{CSC} is depicted in \figurename~\ref{fig:feelBlockDiagram}.

The \ac{CP} duration is assumed to be larger than the maximum-excess delays of the channels between the \acp{ED} and \ac{ES}. Thus, the $\indexOFDMSymbol$th received baseband signal in discrete-time can be written as
\begin{align}
    \receivedVector[\indexOFDMSymbol] =\sum_{\indexED=0}^{\numberOfEdgeDevices-1}\sqrt{\powerED[{\distanceED[\indexED]}]}\channelMatrix[\indexED]\transmittedVector[\indexED,\indexOFDMSymbol]+\noiseVector[\indexOFDMSymbol]~,
\end{align}
where $\channelMatrix[\indexED]\in\complexNumbers^{\idftSize\times\idftSize}$ is a circular-convolution matrix based on the \ac{DFT} of the \ac{CIR} of the channels between \acp{ED} and \ac{ES}, and $\noiseVector[\indexOFDMSymbol]\sim\mathcal{CN}(\zeroVector[{\idftSize}],\noiseVariance\identiyVector[{\idftSize}])$ is the \ac{AWGN}. At the \ac{ES}, the aggregated symbols on the bins can be expressed as
\begin{align}
\symbolVectorEstimate[\indexOFDMSymbol] =\transformDecoder[\numberOfActiveSubcarriers]\frequencyMapping^{\rm H}\diagonalization[{\fdssVector}^{\rm H}]\dftMatrix[\idftSize]\receivedVector[\indexOFDMSymbol]~,
\end{align}
where $\symbolVectorEstimate[\indexOFDMSymbol]\in\complexNumbers^{\numberOfActiveSubcarriers}$ are the received symbols on the bins.
 
% Please cite them selectively:
%\cite{Safi_2020_CCNC, Safi_2020_GC, Sahin_DFRC}

\subsection{Learning Model}
Let $\completeData$ be the dataset containing all the labeled data samples. Also, let the vectors $\sampleData[]$ and $\sampleLabel[]$ be a data sample and its associated label, respectively for $\{(\sampleData[],\sampleLabel[])\}\in\completeData$. Let $\dataset[\indexED]$ denote the local dataset for user index, $\indexED=0,1,\dots,\numberOfEdgeDevices-1$ such that $\completeData=\bigcup_{\indexED=1}^{\numberOfEdgeDevices}{\dataset[\indexED]}$. The centralized loss function can be expressed as
\begin{align}
 \lossFunctionGlobal[\modelParameters]&= \frac{1}{|\completeData|}\sum_{\forall(\sampleData[], \sampleLabel[] )\in\completeData} \lossFunctionSample[{\modelParameters,\sampleData[],\sampleLabel[]}]\nonumber\\&=\sum_{\indexED=1}^{\numberOfEdgeDevices}{ \frac{1}{|\dataset[\indexED]|}\sum_{\forall(\sampleData[], \sampleLabel[] )\in\dataset[\indexED]} \lossFunctionSample[{\modelParameters,\sampleData[],\sampleLabel[]}]}=\frac{1}{\numberOfEdgeDevices}\sum_{\indexED=1}^{\numberOfEdgeDevices}{\lossFunctionLocal[\indexED][\modelParameters]}~,
	\label{eq:cenLossFuncGlobal}
\end{align}
where $\modelParameters=[\modelParametersEle[1],\mydots,\modelParametersEle[\numberOfModelParameters]]^{\rm T}\in\realNumbers^{\numberOfModelParameters}$ is the parameter vector,
$\lossFunctionSample[{\modelParameters,\sampleData[],\sampleLabel[]}]$ denotes the sample loss function that measures the labeling error for $(\sampleData[], \sampleLabel[])$. 
% \begin{align}
%  \lossFunctionGlobal[\modelParameters] &=\sum_{\indexED=1}^{\numberOfEdgeDevices}{ \frac{1}{|\dataset[\indexED]|}\sum_{\forall(\sampleData[], \sampleLabel[] )\in\dataset[\indexED]} \lossFunctionSample[{\modelParameters,\sampleData[],\sampleLabel[]}]}=\frac{1}{\numberOfEdgeDevices}\sum_{\indexED=1}^{\numberOfEdgeDevices}{\lossFunctionLocal[\indexED][\modelParameters]}~.
% 	\label{eq:cenLossFuncLoc}
% \end{align}
In the case of distributed learning, the goal is to minimize the loss function in \eqref{eq:cenLossFuncGlobal} to find the desired parameter vector $\modelParameters$, i.e.,
\begin{align}
    \modelParameters^* = \arg \min_{\modelParameters} {\lossFunctionGlobal[\modelParameters]}
    \label{eq:modelParam}~,
\end{align}
where the dataset are not uploaded to a centralized server.

To solve \eqref{eq:modelParam}, the procedure for distributed training by the \ac{MV} based on \ac{signSGD} \cite{Bernstein_2018} can be summarized as follows: Let $\localGradient[\indexED][\indexCommunicationRound]$ be the local stochastic gradient vector of the $\indexED$th \ac{ED}, given by
\begin{align}
\localGradient[\indexED][\indexCommunicationRound] =  \nabla \lossFunctionLocal[\indexED][{\modelParametersAtIteration[\indexCommunicationRound]}] 
= \frac{1}{\batchSize} \sum_{\forall(\sampleData[], \sampleLabel[] )\in\datasetBatch[\indexED]} \nabla 
\lossFunctionSample[{\modelParametersAtIteration[\indexCommunicationRound],\sampleData[],\sampleLabel[]}]
~,
	\label{eq:LocalGradientEstimate}
\end{align}
where $\datasetBatch[\indexED]\subset\dataset[\indexED]$ is the set of the selected data samples and $\batchSize=|\datasetBatch[\indexED]|$ is the batch size, $\modelParametersAtIteration[\indexCommunicationRound]$ is the parameter vector at the $\indexCommunicationRound$th communication round. Let $\localGradientElement[\indexED,\indexGradient][\indexCommunicationRound]$ be the $\indexGradient$th element of $\localGradient[\indexED][\indexCommunicationRound]$. All \acp{ED} calculate their votes as ${\localGradientSignElement[\indexED,\indexGradient][\indexCommunicationRound]}\triangleq {\signNormal[{\localGradientElement[\indexED,\indexGradient][\indexCommunicationRound]}]}, \forall\indexGradient,\indexED$, and provide them to the \ac{ES}. The \ac{ES} then obtains the \ac{MV} $\majorityVoteEle[\indexCommunicationRound][\indexGradient]$ for the $\indexGradient$th gradient with the expression given by
\begin{align}
\majorityVoteEle[\indexCommunicationRound][\indexGradient]\triangleq\sign\left(\sum_{\indexED=1}^{\numberOfEdgeDevices} {\signNormal[{\localGradientElement[\indexED,\indexGradient][\indexCommunicationRound]}]}\right)=\sign\left(\sum_{\indexED=1}^{\numberOfEdgeDevices} {\localGradientSignElement[\indexED,\indexGradient][\indexCommunicationRound]}\right), \forall\indexGradient~.
\label{eq:majorityVote}
\end{align}
Afterwards, the \ac{ES} sends the calculated \ac{MV} vector $\majorityVote[\indexCommunicationRound]\triangleq[\majorityVoteEle[\indexCommunicationRound][1],\mydots,\majorityVoteEle[\indexCommunicationRound][\numberOfModelParameters]]^{\rm T}$ to the \acp{ED}. All the \acp{ED} update their parameters for the next communication round as
\begin{align}
\modelParametersAtIteration[\indexCommunicationRound+1] = \modelParametersAtIteration[\indexCommunicationRound] - \learningRate \majorityVote[\indexCommunicationRound]~,
\label{eq:MVupdate}
\end{align}
where $\learningRate$ is the learning rate. 

In this study, we consider the same procedure outlined above. However, we consider its implementation in a wireless network and calculate the \ac{MV} in \eqref{eq:majorityVote} with an \ac{OAC} scheme that relies on \acp{CSC}.

\section{CSC-Based Majority Vote}
\label{sec:ppmMV}
In this section, we discuss the transmitter at \acp{ED} and the receiver \ac{ES} with the proposed \ac{OAC} scheme. We also elaborate on the convergence rate of the distributed learning when the proposed \ac{OAC} is used for the \ac{MV} calculation.

\subsection{Edge Device - Transmitter}
%We consider the mapping rule defined in \cite{sahinWCNC_2022} as our scheme is compatible with \ac{DFT-s-OFDM}. 
We encode the gradients with \acp{CSC} as follows: Let $\numberOfOFDMSymbols$ be the number of blocks used in the transmission to train the model for each communication round. Each block can synthesize $\numberOfActiveSubcarriers$ \acp{CSC} corresponding to the $\numberOfActiveSubcarriers$ \ac{DFT-s-OFDM} bins. From the $\numberOfActiveSubcarriers\numberOfOFDMSymbols$ available indices, we propose to assign two active indices for the $\indexGradient$ th gradient. In this case, the number of votes that can be carried for each block is equal to $    \numberOfVotesPerDFTsOFDM = \floor*{{\numberOfActiveSubcarriers}/{(2+2\uniformGap)}}$. To express the mapping rigorously, let $\mappingFunction$ be a pre-defined function that maps $\indexGradient\in\{1,2,\mydots,\numberOfModelParameters\}$ to the distinct pairs $(\voteInTime[+],\voteInFrequency[+])$ and $(\voteInTime[-],\voteInFrequency[-])$ for $\voteInTime[+],\voteInTime[-]\in\{0,1,\mydots,\numberOfOFDMSymbols-1\}$ and  $\voteInFrequency[+],\voteInFrequency[-]\in\{0,1,\mydots,2\numberOfVotesPerDFTsOFDM-1\}$. Let  $(\symbolVector[\indexED,{\indexOFDMSymbol}])_{\indexSubcarrier}$ be the symbol at the $\indexSubcarrier$th index of the $\indexOFDMSymbol$th symbol for the $\indexED$th user. For all $\indexGradient$, 
 we then set the corresponding symbols as
%  $(\symbolVector[\indexED,{\voteInTime[+]}])_{\voteInFrequency[+](1+\uniformGap)} =  \randomSymbolAtSubcarrier[\indexED,\indexGradient]\indicatorFunction[{\localGradientSignElement[\indexED,\indexGradient][\indexCommunicationRound] =1}]$ and $(\symbolVector[\indexED,{\voteInTime[-]}])_{\voteInFrequency[-](1+\uniformGap)} = \randomSymbolAtSubcarrier[\indexED,\indexGradient]\indicatorFunction[{\localGradientSignElement[\indexED,\indexGradient][\indexCommunicationRound] =-1}]$ 
% \vspace{-2mm}
\begin{align}
	(\symbolVector[\indexED,{\voteInTime[+]}])_{\voteInFrequency[+](1+\uniformGap)} =  \randomSymbolAtSubcarrier[\indexED,\indexGradient]\indicatorFunction[{\localGradientSignElement[\indexED,\indexGradient][\indexCommunicationRound] =1}]~,
\end{align}
% \vspace{-2mm}
and
\begin{align}
	(\symbolVector[\indexED,{\voteInTime[-]}])_{\voteInFrequency[-](1+\uniformGap)} = \randomSymbolAtSubcarrier[\indexED,\indexGradient]\indicatorFunction[{\localGradientSignElement[\indexED,\indexGradient][\indexCommunicationRound] =-1}]~,
\end{align}
where $\randomSymbolAtSubcarrier[\indexED,\indexGradient]$ is a random symbol on the unit circle to introduce randomness to the synthesized symbols and $\uniformGap\ge0$ is a parameter to set the guard period between the \acp{CSC}. Due to the channel dispersion, delay spread, and potential time synchronization errors, the \acp{CSC} can interfere with each other. We deactivate $\uniformGap$ indices followed by any active index to avoid interference. The inactivated indices provide a guard period of $\guardTimeUniform = \uniformGap\symbolDuration/\numberOfActiveSubcarriers$ between two adjacent \acp{CSC}. Let the delay due to time-synchronization error be $\syncError$~seconds and the maximum delay of the channel be $\channelSpread$~seconds. Hence, the negligible interference can be ensured by choosing $\uniformGap$ under the condition given by
% \vspace{-1mm}
\begin{align}
\guardTimeUniform=\frac{\uniformGap\symbolDuration}{\numberOfActiveSubcarriers}\ge\channelSpread+\syncError~.
\label{eq:condition}
\end{align}
\vspace{0.2mm}
%  only for patent
\subsection{Edge Server - Receiver}
The mapping function $\mappingFunction$ is assumed to be known to the \ac{ES} so that the \ac{ES} can calculate the pairs $(\voteInTime[+],\voteInFrequency[+])$ and $(\voteInTime[-],\voteInFrequency[-])$ for a given $\indexGradient$. The \ac{MV} for the $\indexGradient$th gradient can be obtained as
\begin{align}
	\majorityVoteEle[\indexCommunicationRound][\indexGradient] = \signNormal[{\deltaVectorAtIteration[\indexCommunicationRound][\indexGradient]}]~,
	\label{eq:detector}
\end{align}
where $\deltaVectorAtIteration[\indexCommunicationRound][\indexGradient]\triangleq{\metricForFirst[\indexGradient]-\metricForSecond[\indexGradient]}$ for $ \metricForFirst[\indexGradient]\triangleq \sum_{\indexSubcarrier=\voteInFrequency[+](1+\uniformGap)}^{(\voteInFrequency[+]+1)(1+\uniformGap)-1}{|{(\symbolVectorEstimate[{\voteInTime[+]}])_{\indexSubcarrier}}|^2}$ and $\metricForSecond[\indexGradient] \triangleq \sum_{\indexSubcarrier=\voteInFrequency[-](1+\uniformGap)}^{(\voteInFrequency[-]+1)(1+\uniformGap)-1}{|{(\symbolVectorEstimate[{\voteInTime[-]}])_{\indexSubcarrier}}|^2}$. We consider $(1+\uniformGap)$ bins for the energy calculations for $\metricForFirst[\indexGradient]$ and $\metricForSecond[\indexGradient]$. The transmitter and the receiver block diagrams are provided in \figurename~\ref{fig:feelBlockDiagram}.

\subsection{Convergence Rate}
The \ac{MV} computed with \eqref{eq:detector} obtains the original \ac{MV} given in \eqref{eq:majorityVote}, probabilistically, due to the non-coherent detection. Nevertheless, for a non-convex loss function $\lossFunctionGlobal[\modelParameters]$, we can show that \ac{CSC-MV} still maintains the convergence of the original \ac{MV} in \cite{Bernstein_2018} under the assumptions given as follows:
% \begin{align}
% 	\metricForFirst[\indexGradient]\triangleq  \sum_{\indexSubcarrier=\voteInFrequency[+](1+\uniformGap)}^{(\voteInFrequency[+]+1)(1+\uniformGap)-1}{|{(\symbolVectorEstimate[{\voteInTime[+]}])_{\indexSubcarrier}}|^2}~,
% 	\label{eq:energyFirst}
% \end{align}
% and
% \begin{align}
% 	\metricForSecond[\indexGradient] \triangleq  \sum_{\indexSubcarrier=\voteInFrequency[-](1+\uniformGap)}^{(\voteInFrequency[-]+1)(1+\uniformGap)-1}{|{(\symbolVectorEstimate[{\voteInTime[-]}])_{\indexSubcarrier}}|^2}~.
% 	\label{eq:energySecond}
% \end{align}
% Due to the delay spread and the synchronization errors, the position of the chirps in the time domain can be changed at the receiver side. To tackle the problem, we consider $(1+\uniformGap)$ bins for the energy calculations in \eqref{eq:energyFirst} and \eqref{eq:energySecond}. The transmitter and the receiver block diagrams are provided in \figurename~\ref{fig:feelBlockDiagram}.
% The \ac{MV} obtained from \eqref{eq:detector} probabilisticaly approximates the original \ac{MV} given in \eqref{eq:majorityVote}. For a non-convex loss function $\lossFunctionGlobal[\modelParameters]$, based on the assumptions stated in \cite{sahinWCNC_2022}, we can show that our proposed scheme maintain the convergence of the original \ac{MV} in \cite{Bernstein_2018}. %However the convergence analysis is not provided in this paper for limitations.

\begin{assumption}[Bounded loss function]
	\rm 
	$\forall\modelParameters$, $\exists\lossFunctionGlobalMinimum$ such that $\lossFunctionGlobal[\modelParameters]\ge \lossFunctionGlobalMinimum$.
\end{assumption}
\begin{assumption}[$L$-smooth gradient \cite{nesterov2004introductory}] 
	\rm 
	Let $\globalGradientNoIndex$ be the gradient of $\lossFunctionGlobal[\modelParameters]$ evaluated at $\modelParameters$. $\forall\modelParameters$ and $\forall\modelParameters'$, the expression given by
	\begin{align}
		\left| \lossFunctionGlobal[\modelParameters'] - (\lossFunctionGlobal[\modelParameters]-\globalGradientNoIndex^{\rm T}(\modelParameters'-\modelParameters)) \right| \le \frac{1}{2}\sum_{\indexGradient=1}^{\numberOfModelParameters} \nonnegativeConstantsEle[\indexGradient](\modelParametersEle[\indexGradient]'-\modelParametersEle[\indexGradient])^2~,
		\nonumber
	\end{align}	
	holds for some vector with non-negative constant values, i.e., $\nonnegativeConstants=[\nonnegativeConstantsEle[1],\mydots,\nonnegativeConstantsEle[\numberOfModelParameters]]^{\rm T}$.
\end{assumption}
\begin{assumption}[Bounded variance]
	\rm The local estimates of the stochastic gradient, $\{\localGradientNoIndex[\indexED]=[\localGradientNoIndexElement[\indexED,1],\mydots,\localGradientNoIndexElement[\indexED,\numberOfModelParameters]]^{\rm T}=\nabla \lossFunctionLocal[\indexED][{\modelParametersAtIteration[\indexCommunicationRound]}]\} $, $\forall\indexED$, are independent and unbiased estimates of $\globalGradientNoIndex=[\globalGradientElementNoIndex[1],\mydots,\globalGradientElementNoIndex[\numberOfModelParameters]]^{\rm T}=\nabla\lossFunctionGlobal[{\modelParameters}]$ with a coordinate bounded variance, i.e., $ \expectationOperator[{\localGradientNoIndex[\indexED]}][]=\globalGradientNoIndex,~\forall\indexED$ and $\expectationOperator[{(\localGradientNoIndexElement[\indexED,\indexGradient]-\globalGradientElementNoIndex[\indexGradient])^2}][]\le\varianceBoundEle[\indexGradient]^2/\batchSize,~\forall\indexED,\indexGradient$, 	where $\varianceBound = [\varianceBoundEle[1],\mydots,\varianceBoundEle[\numberOfModelParameters]]^{\rm T}$ is a non-negative constant vector.
% 	\begin{align}
% 		\expectationOperator[{\localGradientNoIndex[\indexED]}][]&=\globalGradientNoIndex,~\forall\indexED,\\	\expectationOperator[{(\localGradientNoIndexElement[\indexED,\indexGradient]-\globalGradientElementNoIndex[\indexGradient])^2}][]&\le\varianceBoundEle[\indexGradient]^2/\batchSize,~\forall\indexED,\indexGradient,
% 	\end{align}
\end{assumption}
\begin{assumption}[Unimodal, symmetric gradient noise]
	\rm
	For a given $\modelParameters$, each element of the vector $\localGradientNoIndex[\indexED]$, $\forall\indexED$, follows a unimodal distribution that is symmetric around its mean.
\end{assumption}
We assume that \acp{CSC} are orthogonal to each other. This assumption is not strong because the interference among them can be maintained negligibly low under \eqref{eq:condition}. Hence, by using the steps mentioned in \cite{sahinWCNC_2022}, based on the aforementioned assumptions, the following theorem can be derived:
\begin{theorem}
\rm For $\batchSize=\communicationRounds/\brachSizeRelativeToRounds$ and $\learningRate=1/\sqrt{\norm{\nonnegativeConstants}_1\batchSize}$, the convergence rate of \ac{CSC-MV} based \ac{FEEL} in fading channel can be expressed as,
\begin{align}
\expectationOperator[\frac{1}{\communicationRounds}\sum_{\indexCommunicationRound=0}^{\communicationRounds-1} \norm{\globalGradient[\indexCommunicationRound]}_1][]\nonumber\le\frac{1}{\sqrt{\communicationRounds}}&\bigg( \coefficientOne\sqrt{\norm{\nonnegativeConstants}_1}\left(	\lossFunctionGlobal[{\modelParametersAtIteration[0]}]- \lossFunctionGlobalMinimum+\frac{\brachSizeRelativeToRounds}{2}\right)\nonumber\\&~~+\frac{2\sqrt{2\brachSizeRelativeToRounds}}{3}\norm{\varianceBound}_1\bigg)~,
\label{eq:convergence}
\end{align}
where $\coefficientOne=(1+\frac{2}{\effectiveSNR\numberOfEdgeDevices})\frac{1}{\sqrt{\brachSizeRelativeToRounds}}$ for $\effectiveSNR\triangleq\frac{\symbolEnergy}{(1 + \uniformGap)\noiseVariance}$.
\label{th:convergence}
\end{theorem}

\subsection{Trade-offs and Comparisons}
\label{subsec:tradeoff}
The schemes in \cite{Guangxu_2020, Guangxu_2021, Amiria_2021} rely on \ac{OFDM} modulation. However, \ac{OFDM} is known to suffer from high \ac{PMEPR}.
The main advantage of \ac{CSC-MV} is that it can achieve a significantly low \ac{PMEPR} compared to the methods in \cite{Guangxu_2020, Guangxu_2021, Amiria_2021, sahinWCNC_2022}. The scheme in \cite{sahinWCNC_2022} is also based on \ac{DFT-s-OFDM} leading to low \ac{PMEPR} symbols. However, theoretically, \ac{CSC-MV} can decrease the \ac{PMEPR} even further and can achieve a \ac{PMEPR} as low as $0~$dB (for $\numberOfVotesPerDFTsOFDM=1$), unlike the method in \cite{sahinWCNC_2022}\footnote{Practically, the \ac{PMEPR} is higher than the theoretical value as the \ac{CSC} is distorted due to the truncation of the \ac{FDSS} vector. However, it can be mitigated by either decreasing the frequency deviation of the \ac{CSC} or allowing an amount of leakage beyond the assigned bandwidth \cite{Sahin_CSC2020}.}. For our scheme, the \ac{PMEPR} increases with the number of chirps \cite{Safi_2020_CCNC}. Hence, for $\numberOfVotesPerDFTsOFDM$ votes, the \ac{PMEPR} is $10\log_{10}{\numberOfVotesPerDFTsOFDM}$. The number of symbols needed to train each round is $\ceil{\frac{\numberOfModelParameters}{\numberOfVotesPerDFTsOFDM}}{}$. Thus, \ac{CSC-MV} causes a trade-off between \ac{PMEPR} and resource consumption. As a result, we must increase the \ac{PMEPR} limit to reduce resource utilization.

\ac{OBDA} \cite{Guangxu_2021} and \ac{BAA} \cite{Guangxu_2020} rely on the \ac{CSI} being available to the \acp{ED}. Although the proposed scheme in \cite{Amiria_2021} does not require the \ac{CSI} to be available on the \acp{ED}, the \ac{ES} utilizes multiple antennas to overcome the impact of fading, and the sum of the channel gains from the \ac{ED} to each antenna is assumed to be available to the \ac{ES}. \Ac{CSC-MV} does not rely on the summed \ac{CSI} or multiple antennas. Non-coherent detection also aids in the elimination of synchronization issues as it does not need phase synchronization.

Finally, the methods in \cite{Guangxu_2020} and \cite{Amiria_2021} rely on analog transmission, which is incompatible with the existing digital modulation-based wireless systems. Similar to \ac{OBDA} \cite{Guangxu_2021}, \Ac{CSC-MV} is compatible with digital modulation. Also, since the proposed scheme synthesizes \ac{CSC} over \ac{DFT-s-OFDM}, it is  compatible with the transceivers in \ac{5G} \ac{NR}.
% The non-lthe \ac{CSC-MV} needs $\ceil{\frac{\numberOfModelParameters}{\numberOfVotesPerDFTsOFDM}}{}$ symbols to train each round, whereas the \ac{PMEPR} would be $10\log{\numberOfVotesPerDFTsOFDM}$. As a result, we must increase the \ac{PMEPR} limit to reduce resource utilization.inear distortion depends on the \ac{PMEPR} of the transmitted waveform. When a high \ac{PMEPR} signal passes through a power amplifier, the amount of distortion is higher.

\begin{figure}[t]
	\centering
	{\includegraphics[width =3.2in]{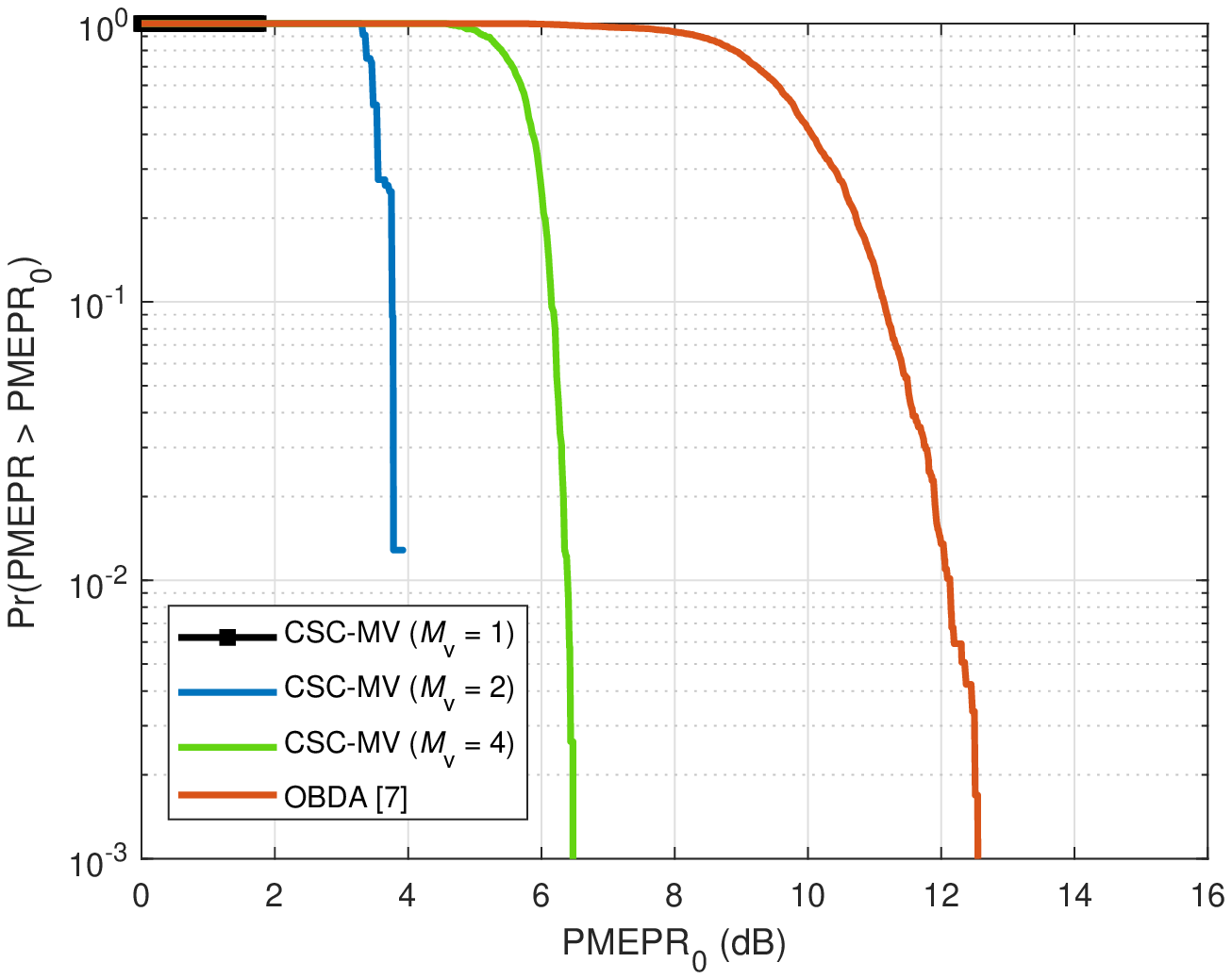}
	} 
	\caption{PMEPR distributions.}
	\label{fig:pmepr}
		\centering
	{\includegraphics[width =3.2in]{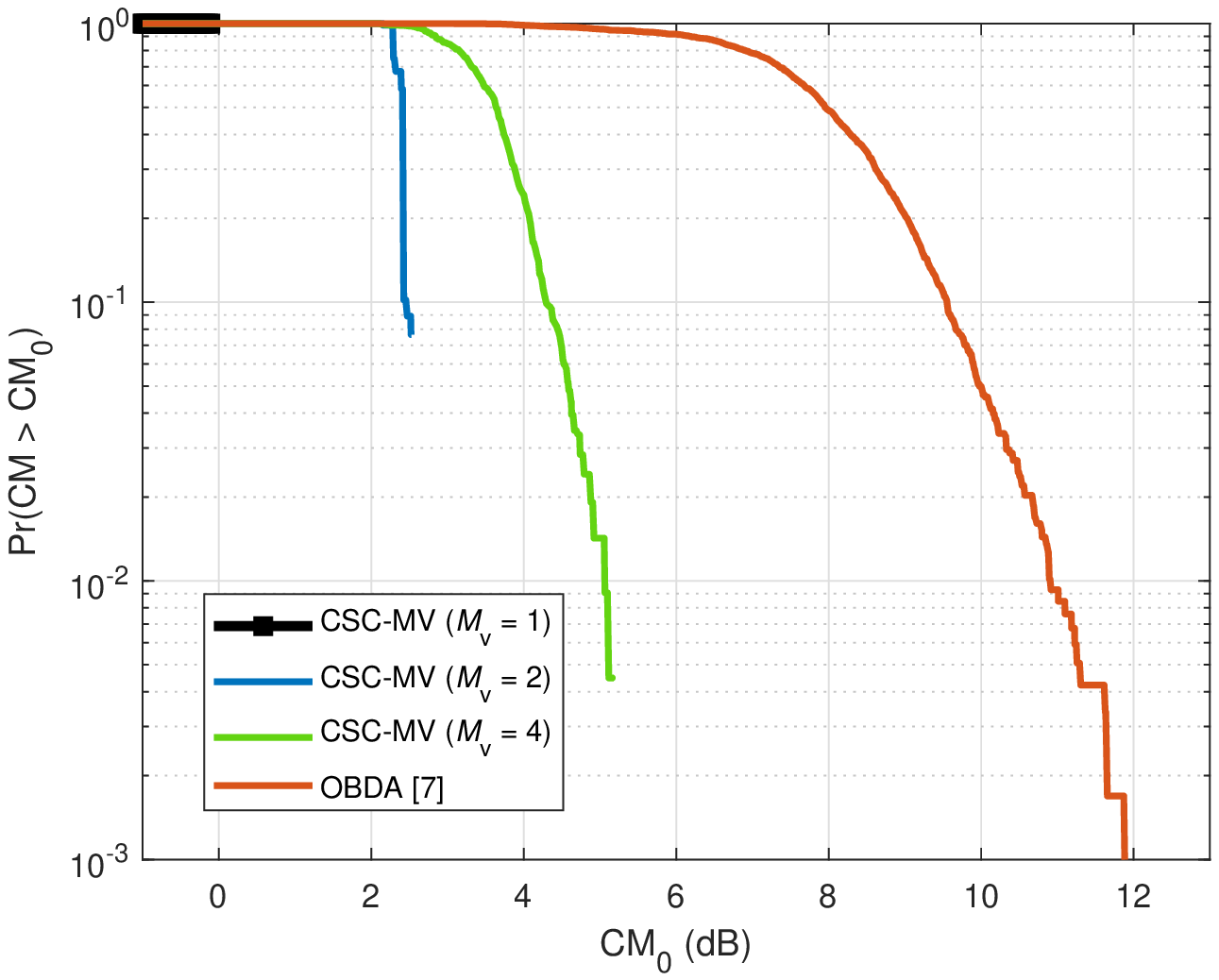}
	} 
	\caption{CM distributions.}
	\label{fig:cm}
%\end{figure}
%\begin{figure}[t]
	\centering
	{\includegraphics[width= \figuresizeWidth]{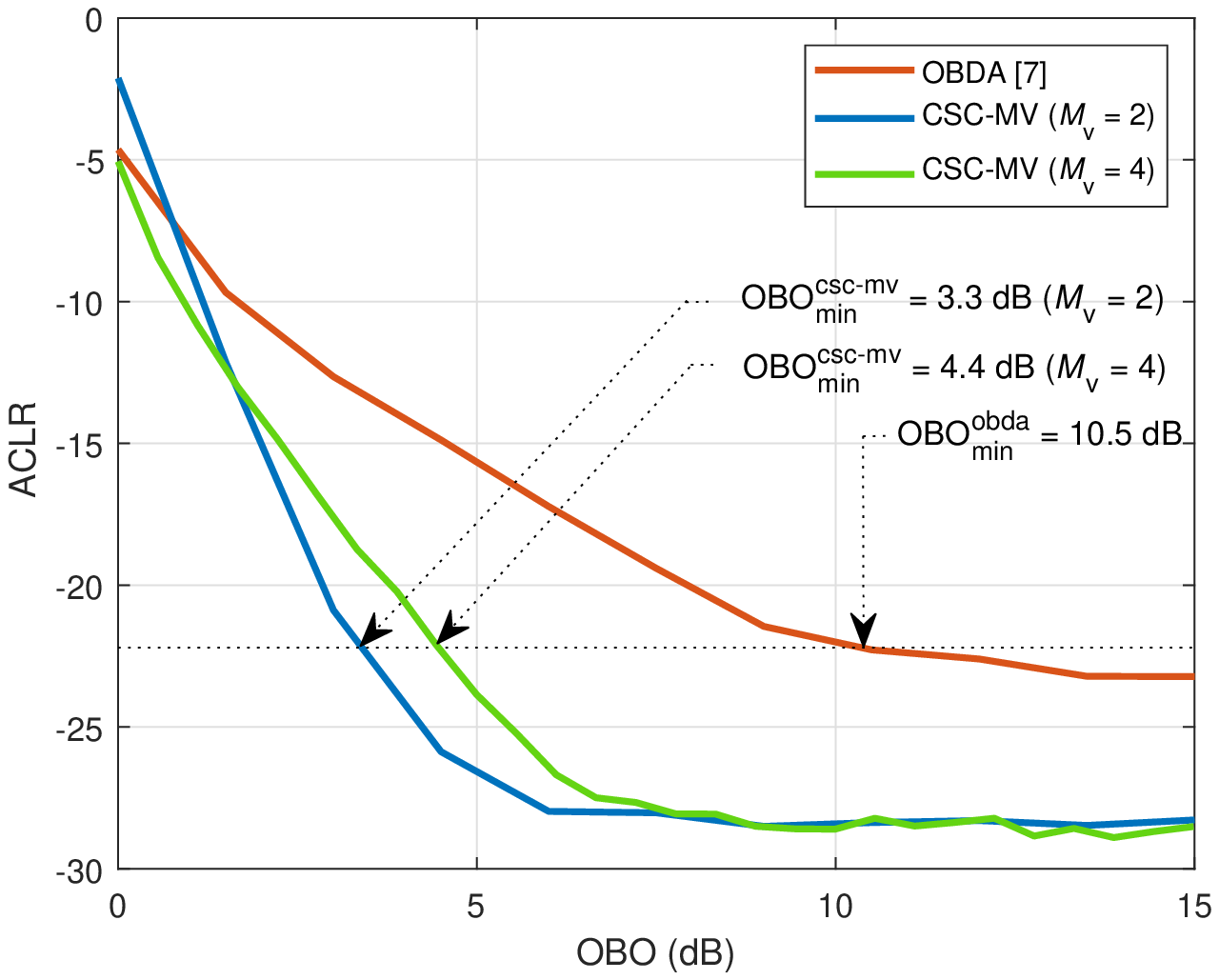}
	} 
	\caption{\ac{ACLR} versus \ac{OBO} plot for different schemes.}
	\label{fig:obovsaclr}
\end{figure}

\begin{figure}[t]
\centering
	\subfloat[Operating at $\oboVar=10.5$~dB.]{\includegraphics[width =\figuresizeWidth]{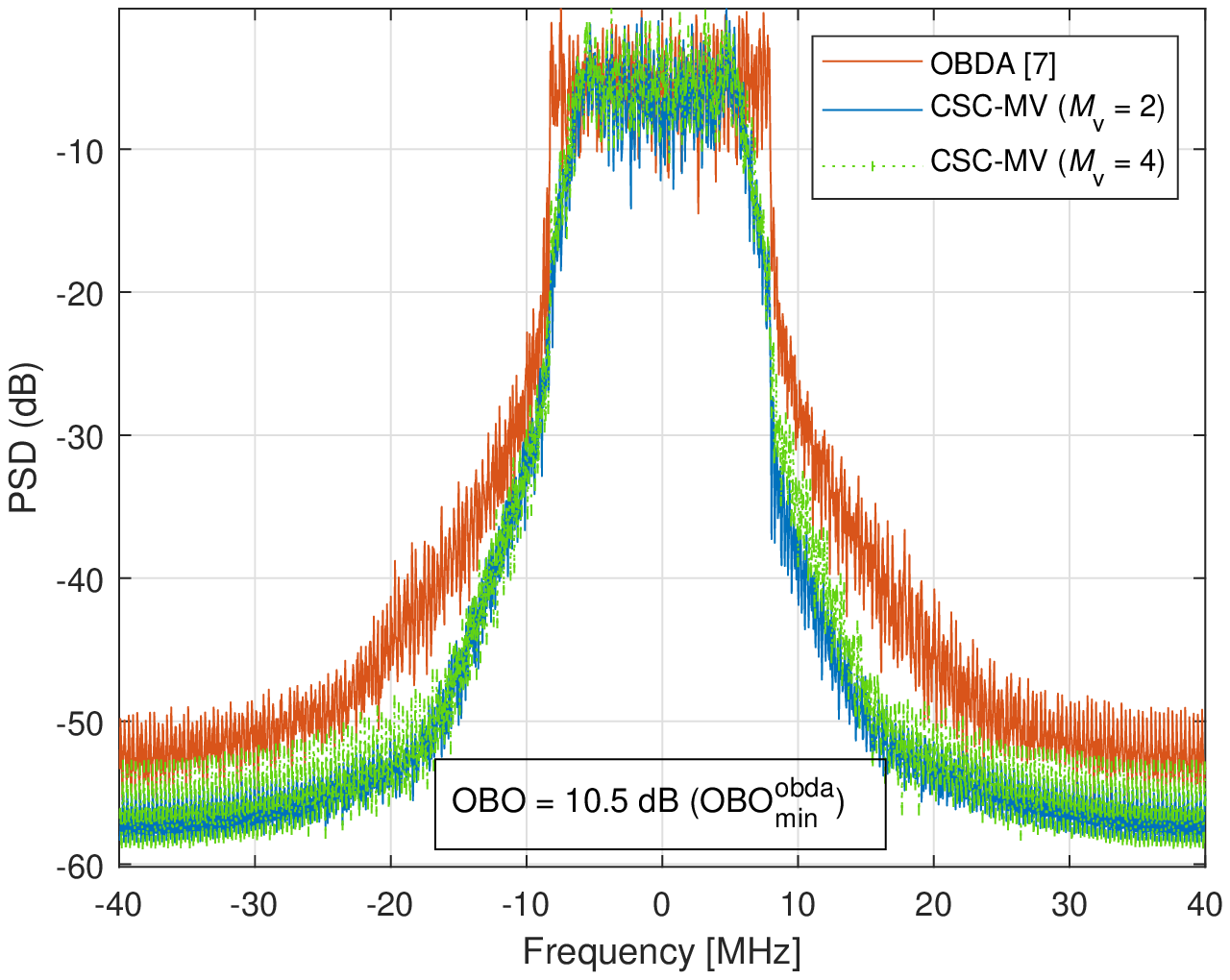}
		\label{subfig:oobchirp}}~\\
	\subfloat[Operating at $\oboVar=3.3$~dB.]{\includegraphics[width =\figuresizeWidth]{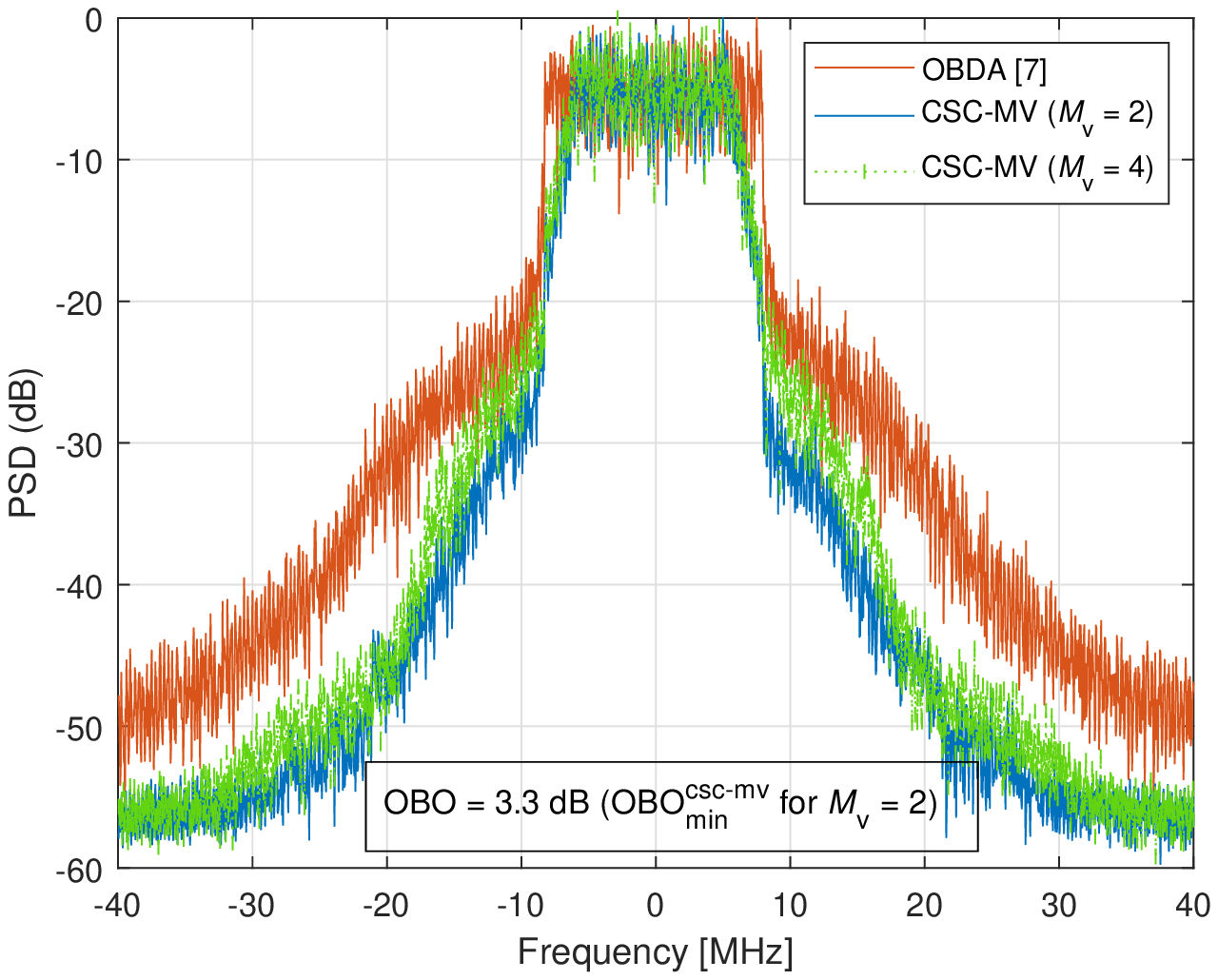}
		\label{subfig:oobobda}}~
	\caption{\ac{OOB} performance of \ac{CSC-MV} and \ac{OBDA} at different \ac{OBO} levels. }
	\label{fig:oob}
	\centering
	{\includegraphics[width =3in]{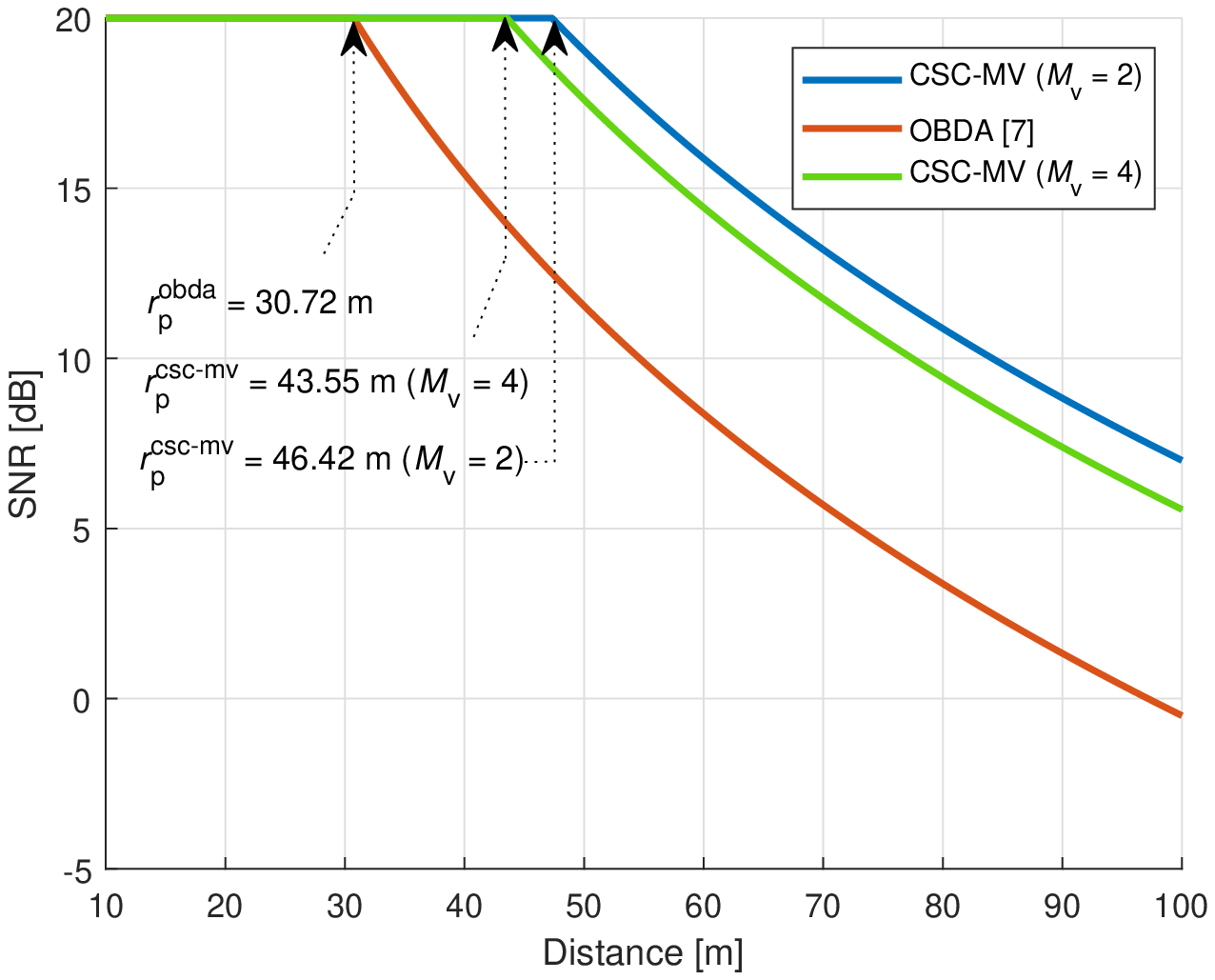}
	} 
	\caption{\ac{SNR} versus link distance performance for $\oboMax= 30$~dB.}
	\label{fig:snrvsdistance}
\end{figure}
\begin{figure}
	\centering
	{\includegraphics[width =3.2in]{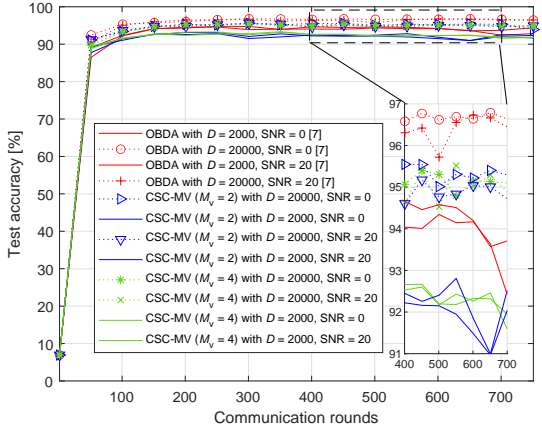}
	} 
	\caption{Test accuracy results for homogeneous data distributions.}
	\label{fig:plotIID}
\end{figure}
\begin{figure*}
	\centering
	\subfloat[SNR is $0$~dB.]{\includegraphics[width =\figuresizeWidth]{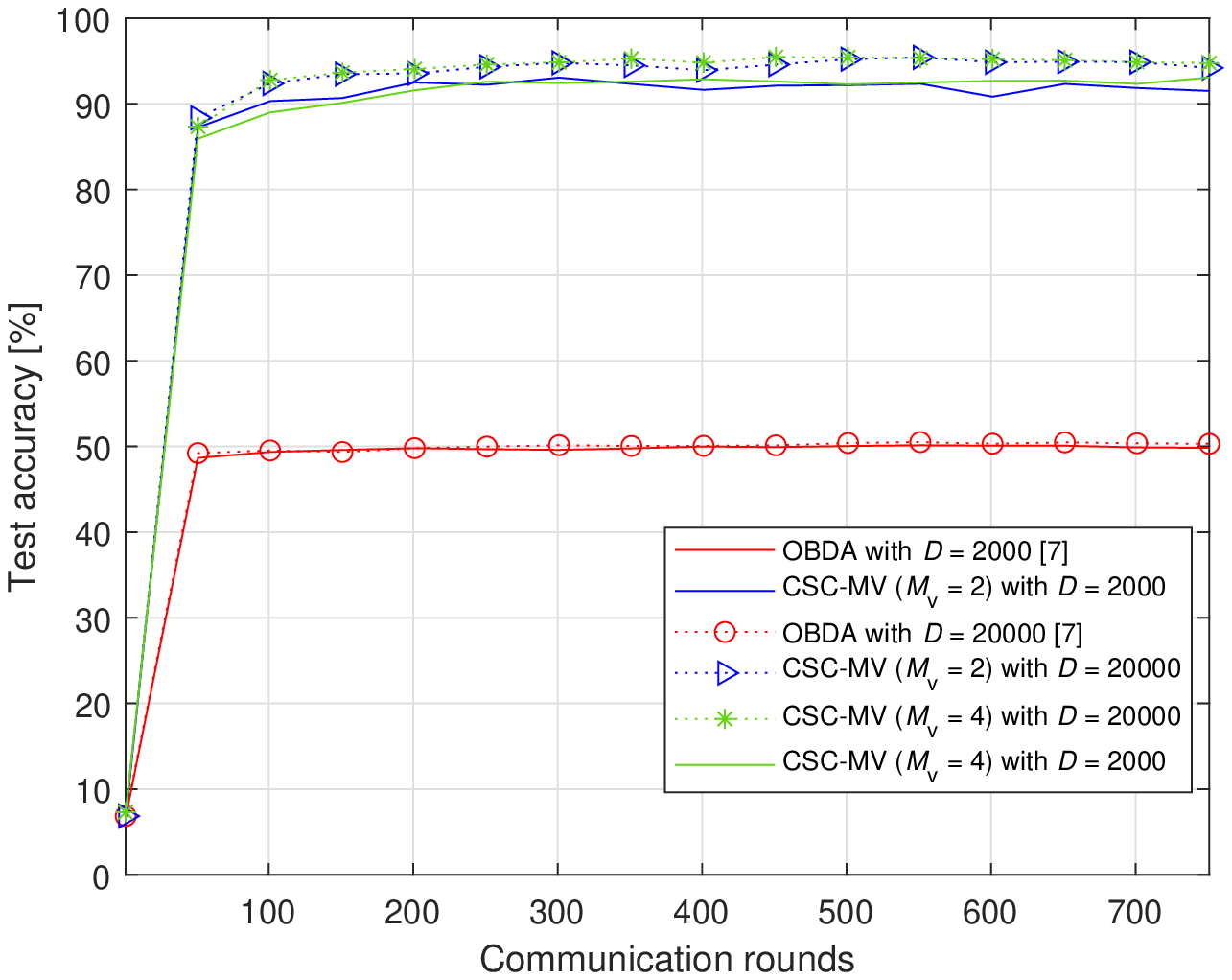}
		\label{subfig:acc0db50eds}}~
	\subfloat[SNR is $20$~dB.]{\includegraphics[width =\figuresizeWidth]{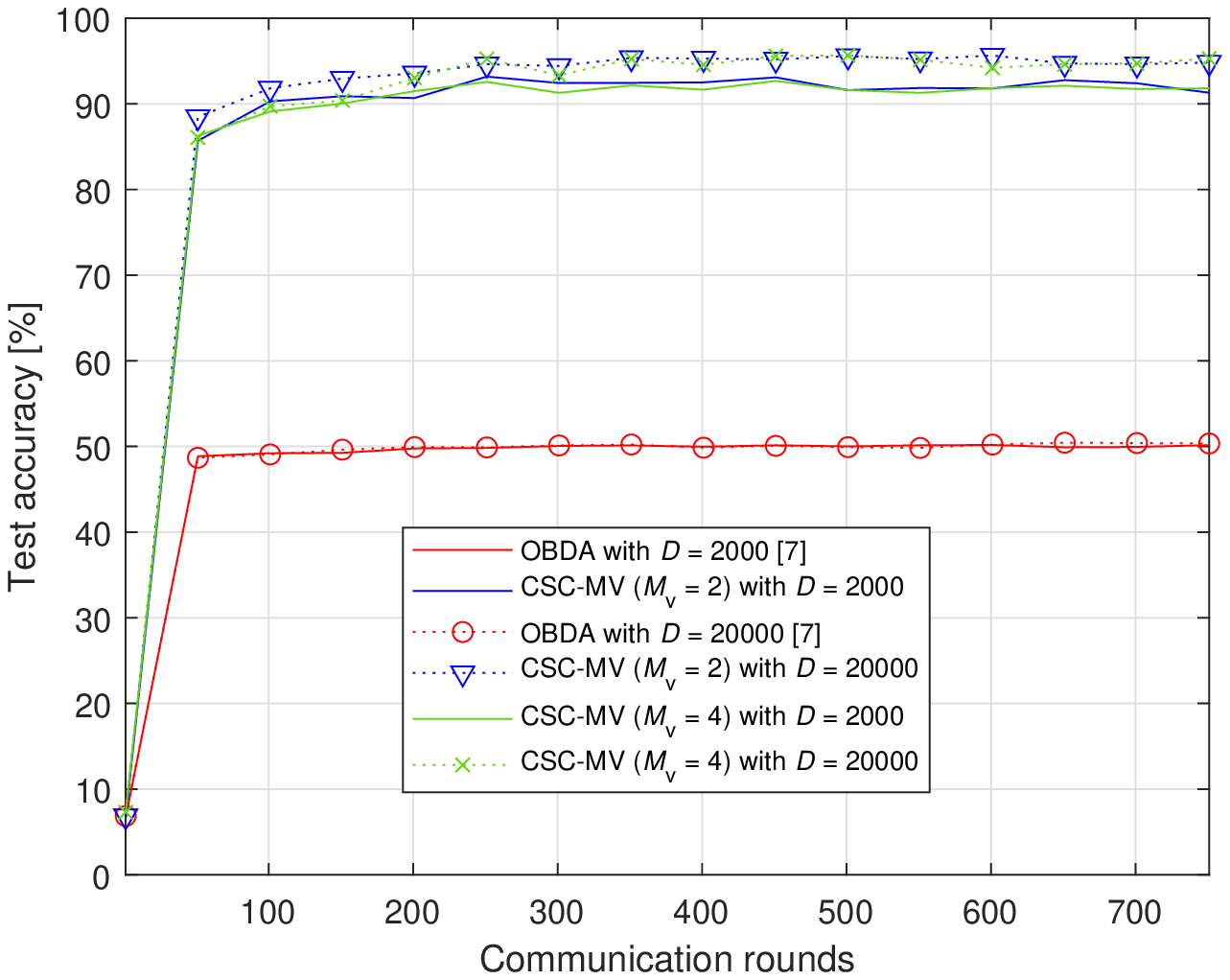}
		\label{subfig:acc20db50eds}}~
	\caption{Average test accuracy results for heterogeneous data distributions (EPA, $\communicationRounds=750$, $\numberOfEdgeDevices=50$).}
	\label{fig:testAccnonIID}
%\end{figure*}
%\begin{figure*}
	\centering
	\subfloat[SNR is $0$~dB.]{\includegraphics[width =\figuresizeWidth]{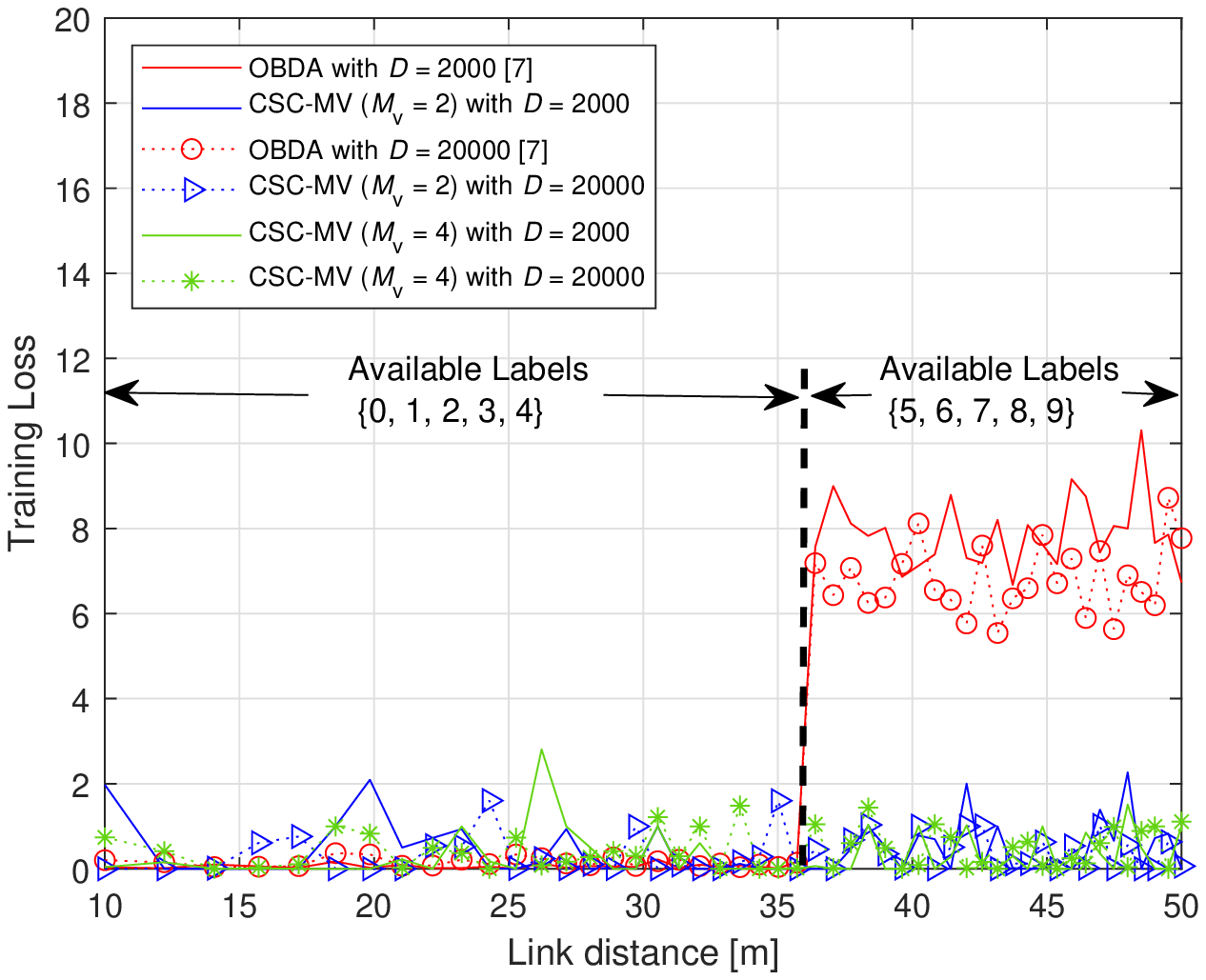}
		\label{subfig:lossDist0db50eds}}~
	\subfloat[SNR is $20$~dB.]{\includegraphics[width =\figuresizeWidth]{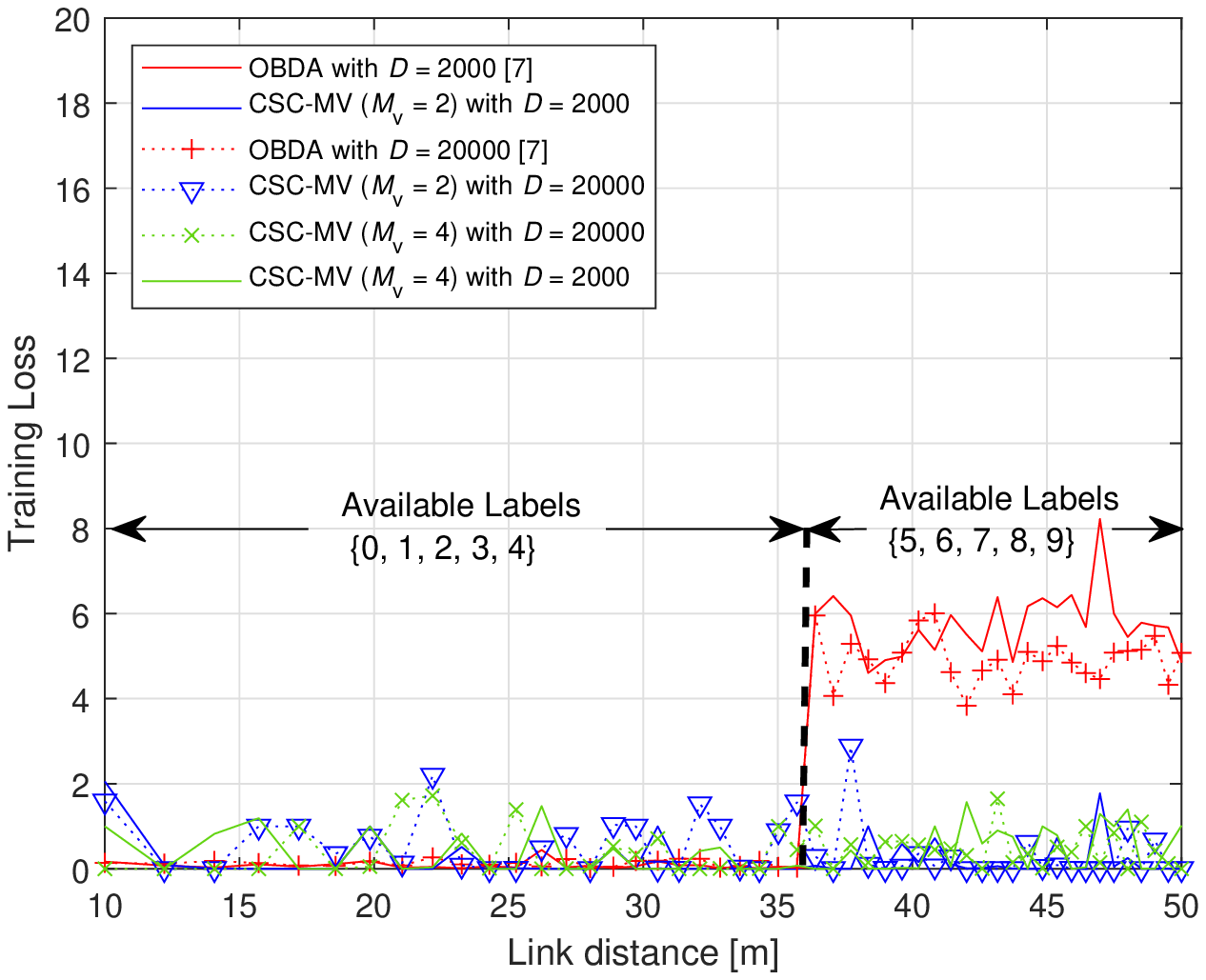}
		\label{subfig:lossDist20db50eds}}~
	\caption{Training loss versus link distance results for heterogeneous data distributions after 750 communication round (EPA, $\numberOfEdgeDevices=50$).}
	\label{fig:lossDist}
\end{figure*}
\section{Numerical Results}
\label{sec:numerical}
We consider the learning task of handwritten digit recognition over a \ac{FEEL} system in a circular cell with a radius of $\cellRadius=50$~m and the number of \acp{ED}, $\numberOfEdgeDevices=50$. The path loss exponent is $\pathLossExpo=4$. We assume perfect power control within the coverage range (i.e., ${\distanceED[\indexED]}<\rangePowercontrol{}$) and $\oboMax$ is set to $30$~dB, and $\referenceDistance = \minimumDistance = 10$~m. We consider the Rapp model for the \ac{PA} at the \acp{ED} with the saturation amplitude of $1$ and the smoothness factor of $3$. We compare the performance of \ac{CSC-MV} with \ac{OBDA} in this setup for both homogeneous and heterogeneous data distributions. For homogeneous data distribution, all digits are equally assigned to each \ac{ED}. For heterogeneous data distributions, the cell is divided into two equal areas with an equal number of \acp{ED}. The first area is the circle with a radius of $\cellRadius/\sqrt{2}$. The second area is the ring-shaped area enclosed by two concentric circles with radius $\cellRadius/\sqrt{2}$ and $\cellRadius$. The \acp{ED} located at the first and the second area only have the data samples with labels $\{0,1,2,3,4\}$ and $\{5,6,7,8,9\}$, respectively (See \cite[Figure 3]{sahinCommnet_2021} for illustration).
 
%  \begin{table}[]
% 	\centering
% 	\caption{Neural network at the EDs.}
% 	\begin{tabular}{l|l}
% 		Layer               				& Learnables \\ \hline\hline
% 		Input ($28\times28$ images) & N/A\\\hline
% 		Convolution 2D ($5\times5$, $20$ filters)   	& \begin{tabular}[c]{@{}l@{}}Weights: $5\times5\times1\times20$\\ Bias: $1\times1\times20$ \end{tabular} \\\hline
% 		Batchnorm & \begin{tabular}[c]{@{}l@{}}Offset: $1\times1\times20$\\ Scale: $1\times1\times20$ \end{tabular} \\ \hline
% 		ReLU &  N/A\\\hline
% 		Convolution 2D ($3\times3$, $20$ filters)   	& \begin{tabular}[c]{@{}l@{}}Weights: $3\times3\times20\times20$\\ Bias: $1\times1\times20$ \end{tabular} \\\hline
% 		Batchnorm & \begin{tabular}[c]{@{}l@{}}Offset: $1\times1\times20$\\ Scale: $1\times1\times20$ \end{tabular} \\ \hline
% 		ReLU &  N/A\\		\hline
% 		Convolution 2D ($3\times3$, $20$ filters)    	& \begin{tabular}[c]{@{}l@{}}Weights: $3\times3\times20\times20$\\ Bias: $1\times1\times20$ \end{tabular} \\\hline
% 		Batchnorm & \begin{tabular}[c]{@{}l@{}}Offset: $1\times1\times20$\\ Scale: $1\times1\times20$ \end{tabular} \\ \hline
% 		ReLU &  N/A\\		\hline
% 		Fully-connected layer ($10$  outputs)   					&    \begin{tabular}[c]{@{}l@{}}Weights: $10\times11520$\\ Bias: $10\times1$ \end{tabular}        \\\hline
% 		Softmax &  N/A\\\hline 
% 	\end{tabular}
% 	\label{table:layout}
% \end{table}
Our model is based on the \ac{CNN} given in \cite{sahinCommnet_2021}, which contains $\numberOfModelParameters=123090$ learnable parameters. We considered $\numberOfActiveSubcarriers=54$ subcarriers and \ac{IDFT} size $\idftSize = 64$. The \ac{FEEL} performance is tested under two different uplink \acp{SNR}, i.e., $0$~dB and $20$~dB. ITU Extended Pedestrian A (EPA) is considered for the fading channel with no mobility for each round and the channel variation is considered by regenerating the channel at each communication round. The \ac{RMS} delay spread of the EPA channel is $\rmsDelaySpread=43.1$~ns. For each round, we transmit $61545$ and $30773$ symbols for $\numberOfVotesPerDFTsOFDM=2$ and $\numberOfVotesPerDFTsOFDM=4$, respectively.

In \figurename~\ref{fig:pmepr}, the \ac{PMEPR} distributions are compared. The \ac{OBDA} can cause substantially high \ac{PMEPR} as the signs of the gradients may result in a constructive addition in the time domain. The \acp{CSC}, on the other hand, result in low \ac{PMEPR} as shown in \ref{subsec:tradeoff}. When $\numberOfVotesPerDFTsOFDM$ is $1$, $2$, or $4$, The \ac{PMEPR} is approximately $2$~dB, $3$~dB, and $6$~dB, respectively. However, theoretically, the \ac{PMEPR} should be $0$~dB, $2$~dB, and $4$~dB when $\numberOfVotesPerDFTsOFDM$ is $1$, $2$, or $4$, respectively. The synthesized signal is distorted due to the abrupt frequency change of the linear \acp{CSC} within a symbol duration\cite{Sahin_CSC2020}. As a result, the observed \ac{PMEPR} is larger than the theoretical bound. However, \ac{PMEPR} can be improved by choosing very small $\numberOfOccupiedSubcarriers$ or by modifying the \ac{FDSS} vector $\fdssVector$ \cite{Sahin_CSC2020}.

The \ac{CM} distributions are shown in \figurename~\ref{fig:cm}. The \ac{CM} of a time-domain signal $\timeDomainOFDM[\timeVar]$ is calculated as $\cubicMetric[{\timeDomainOFDM[\timeVar]}] = ({\rawCubicMetric[{\timeDomainOFDM[\timeVar]}]-\rawCubicMetricRef})/{\slopeFactor}$, where the empirical slope factor $\slopeFactor$ is set to $1.52$ for \ac{OFDM} systems and $\rawCubicMetric[{\timeDomainOFDM[\timeVar]}]$ is defined by
$ \rawCubicMetric[{\timeDomainOFDM[\timeVar]}]= 20 \log_{10}\left(\timeDomainOFDMNorm[{\timeVar}]{} \right)$,
where $\timeDomainOFDMNorm[{\timeVar}]$ is the normalized signal. For the reference signal, we set $\rawCubicMetricRef$ to be $1.52$~dB\cite{3gpp_CM}.
For $\numberOfVotesPerDFTsOFDM=1$, \figurename~\ref{fig:cm} demonstrates that \ac{CSC-MV} performs even better than the reference signal. Similar to the \ac{PMEPR} results, \ac{CSC-MV} outperforms \ac{OBDA} for $\numberOfVotesPerDFTsOFDM \in \{1,2,4\}$.

\figurename~\ref{fig:obovsaclr} shows the \ac{ACLR} versus \ac{OBO} plot for \ac{CSC-MV} and \ac{OBDA}. For both schemes, we consider a time-domain windowing with a raised cosine window to minimize spectral leakage. We define \ac{ACLR} as the ratio of the power received outside the allocated frequency band of the channel to the received power on the assigned channel bandwidth. The plots show that under similar \ac{ACLR} constraints, the power amplifier must operate at a larger \ac{OBO} value for the \ac{OBDA} compared to the \ac{CSC-MV}. Moreover, the lowest \ac{ACLR} that \ac{OBDA} and \ac{CSC-MV} can achieve is $-23$~dB and $-28.22$~dB, respectively. If we consider an \ac{ACLR} constraint of $-22$~dB, we calculate $\oboMinObda=10.5$~dB, $\oboMinCsc=3.3$~dB for $\numberOfVotesPerDFTsOFDM=2$ and $\oboMinCsc=4.4$~dB for $\numberOfVotesPerDFTsOFDM=4$ as the minimum \ac{OBO} values at the \acp{PA} for the corresponding schemes. \figurename~\ref{fig:oob}\subref{subfig:oobchirp} and \figurename~\ref{fig:oob}\subref{subfig:oobobda} show the \ac{OOB} performance for different schemes at $\oboVar=\oboMinCsc$ and $\oboVar=\oboMinObda$, respectively. The plots show that the \ac{OBDA} is more prone to the spectral leakage problem than \ac{CSC-MV}. In \figurename~\ref{fig:snrvsdistance}, the uplink \ac{SNR} versus link distance performances are shown under the \ac{ACLR} constraint. The curves indicate that the power control can maintain the uplink \ac{SNR} up to a range of $\rangePowercontrol$, where the range of power control for \ac{OBDA}, \ac{CSC-MV} ($\numberOfVotesPerDFTsOFDM=2$) and \ac{CSC-MV} ($\numberOfVotesPerDFTsOFDM=4$) are $30.72$~m, $43.55$~m, and $46.42$~m, respectively. Hence, the area of the cell is appropriately doubled with $\numberOfVotesPerDFTsOFDM=2$ as compared to \ac{OBDA}.

\figurename~\ref{fig:plotIID} shows the test accuracy results for homogeneous data distribution for $\ac{SNR}=\{0,20\}$~dB, $\numberOfEdgeDevices=50$, and $\numbOfTrainingImages=\{2000,20000\}$. Both \ac{OBDA} and \ac{CSC-MV} perform with high accuracy in all scenarios. We define all the \acp{ED} at $\distanceEDVar\leq\rangePowercontrol$ as near \acp{ED} and the ones at $\distanceEDVar>\rangePowercontrol$ as far \acp{ED}. For the given setup, the number of near \acp{ED} for \ac{OBDA}, \ac{CSC-MV} ($\numberOfVotesPerDFTsOFDM=2$) and \ac{CSC-MV} ($\numberOfVotesPerDFTsOFDM=4$) are 20, 37, and 42, respectively. For \ac{OBDA}, the votes of the 20 near \acp{ED} have a stronger impact compared to the random votes of the 30 far \acp{ED} that are affected by the imperfect power control. The training remains unaffected and \ac{OBDA} performs well for homogeneous data distributions.

\figurename~\ref{fig:testAccnonIID} shows that \ac{CSC-MV} performs much better than \ac{OBDA} in terms of test accuracy for heterogeneous data distributions. Although Assumption 3 does not hold in the case of heterogeneous data distribution, the test results are still remarkable for \ac{CSC-MV}. The test accuracy results can be further understood from the loss vs. link-distance performance after 750 iterations given in \figurename~\ref{fig:lossDist}. For \ac{OBDA}, the plot shows that the 20 near \acp{ED} only have half of the available labels. As a result, the trained model fails to distinguish the other half of the digits and the test accuracy is near $50\%$, i.e., the learning is biased towards the nearby \acp{ED}. For \ac{CSC-MV} ($\numberOfVotesPerDFTsOFDM=4$), of the 37 near \acp{ED}, 25 of them have the dataset with labels $\{0,1,2,3,4,\}$, and the remaining 12 \acp{ED} have the dataset with labels $\{5,6,7,8,9\}$. The availability of all the labels allows the model to converge with high test accuracy. For the same reason, the test accuracy is high for \ac{CSC-MV} ($\numberOfVotesPerDFTsOFDM=2$).

% ou i0[7i] Vo,bg7i=.;[
% fdh}btr]
% \begi]\yhn{figure*}
% 	\centering
% 	\subfloat[EPA, SNR is $0$~dB  ($\communicationRounds=400$, $\numberOfEdgeDevices=50$).]{\includegraphics[width =\figuresizeWidth]{plot_loss_ZerodB_FiftyEDs.eps}
% 		\label{subfig:loss0db50eds}}~
% 	\subfloat[EPA, SNR is $20$~dB  ($\communicationRounds=400$, $\numberOfEdgeDevices=50$).]{\includegraphics[width =\figuresizeWidth]{plot_loss_TwentydB_FiftyEDs.eps}
% 		\label{subfig:loss20db50eds}}~
% 	\caption{Training loss results for non-\ac{iid} data. The \ac{FEEL} with the \ac{OBDA}-\ac{PPM} works without the  \ac{CSI} at the \acp{ED} and \ac{ES}.}
% 	\label{fig:testAcc}
% 	\vspace{-2mm}
% \end{figure*}

\section{Concluding Remarks}
\label{sec:conclusion}
In this study, we propose a \ac{CSC}-based \ac{OAC} for \ac{FEEL}. To implement \ac{FEEL}, all the design challenges for a practical wireless communication system (e.g., \ac{PMEPR}, \ac{CM}, spectral leakage) must be taken into account. The main advantage of \ac{CSC-MV} is the improved power efficiency with low distortion, which results in larger cell size as compared to \ac{OBDA} under an \ac{ACLR} constraint. Also, \ac{CSC-MV} can work at a much lower \ac{OBO} level than the one for \ac{OBDA}. For a distributed learning scenario, a better \ac{PA} efficiency is crucial for reducing the cost of the devices that operate with a low link budget, e.g., \ac{IoT} sensors. On the other hand, \ac{CSC-MV} requires a larger number of symbols as compared to \ac{OBDA}. We demonstrate that a larger cell size helps to achieve a high test accuracy for heterogeneous data distributions, particularly, when the dataset changes based on the location of the devices. Numerical results show that, compared to \ac{OBDA}, \ac{CSC-MV} leads to a larger area in which \ac{ED} can converge, resulting in high test accuracy. 
In the future, the proposed concept will be enhanced to decrease the number of symbols transmitted while maintaining the power-efficiency.
% while the far \acp{ED} fail to converge a high test accuracy with \ac{OBDA} (due to the smaller cell size), the proposed scheme performs well even when the power control is imperfect.
% \acresetall
\bibliographystyle{IEEEtran}
\bibliography{references}

\end{document}